%% file: SpatExtreme.tex
\documentclass[12pt]{article}

\RequirePackage{amsthm,amsmath,amsfonts,amssymb}
\input{commands}

\RequirePackage[colorlinks,citecolor=blue,urlcolor=blue]{hyperref}
\RequirePackage{graphicx}
\usepackage{caption}
\usepackage{subcaption}
\usepackage{algorithm}
\usepackage{algpseudocode}
\usepackage{booktabs}
\usepackage[]{natbib}

\begin{document}
\begin{center}
{\Large  A Deep Learning Synthetic Likelihood Approximation of a Non-stationary Spatial Model for Extreme Streamflow Forecasting}\\\vspace{6pt}
{\large Reetam Majumder\footnote[1]{North Carolina State University} and Brian J. Reich$^1$}\\
\today
\end{center}

\begin{abstract}\noindent
Extreme streamflow is a key indicator of flood risk, and quantifying the changes in its distribution under non-stationary climate conditions is key to mitigating the impact of flooding events. We propose a non-stationary process mixture model (NPMM) for annual streamflow maxima over the central US (CUS) which uses downscaled climate model precipitation projections to forecast extremal streamflow. Spatial dependence for the model is specified as a convex combination of transformed Gaussian and max-stable processes, indexed by a weight parameter which identifies the asymptotic regime of the process. The weight parameter is modeled as a function of the annual precipitation for each of the two hydrologic regions within the CUS, introducing spatio-temporal non-stationarity within the model. The NPMM is flexible with desirable tail dependence properties, but yields an intractable likelihood. To address this, we embed a neural network within a density regression model which is used to learn a synthetic likelihood function using simulations from the NPMM with different parameter settings.  Our model is fitted using observational data for 1972--2021, and inference carried out in a Bayesian framework. The two regions within the CUS are estimated to be in different asymptotic regimes based on the posterior distribution of the weight parameter. Annual streamflow maxima estimates based on global climate models for two representative climate pathway scenarios suggest an overall increase in the frequency and magnitude of extreme streamflow for 2006--2035 compared to the historical period of 1972--2005.

\vspace{12pt}
{\bf Key words:} Deep Learning, Density regression, Max-stable processes, Gaussian process, Vecchia approximation, Climate change.
\end{abstract}




\section{Introduction}
\label{s:intro}
The increase in the frequency of hydroclimatic extreme events in the last few decades has caused devastating economic damage and claimed thousands of human lives \citep{hirabayashi2013global, winsemius2018disaster}. \citep{Winsemius2016} predicted an increase in this cost due to sea level rise and extreme precipitation events brought about by climate change.  Uncertainty in climate change projections, particularly those associated with precipitation \citep{bhowmik2017multivariate}, also results in significant challenges to the design and maintenance of water infrastructure \cite[e.g.,][]{Vahedifard2017,kasler_hecht_2017}. This is further exacerbated by the complexity of flooding events \citep{merz2014floods,condon2015climate,kundzewicz2017differences,franccois2019design} as extremal streamflow, which is a key measure of flood risk, shows spatial clustering \citep{hirsch2012has,MajumderReichShaby2022}. There is therefore a need to account for spatial and temporal variability (i.e., non-stationarity) in extremal streamflow due to precipitation when assessing current and future flood risk \citep{milly2008stationarity,vogel2011nonstationarity,merz2014floods,kundzewicz2014flood,salas2014revisiting,milly2015critiques,vsraj2016influence}. 

A relatively simple approach to projecting flood risk on the basis of extreme streamflow is by the statistical extrapolation of spatiotemporal trends observed in the historical record. Extreme value analysis (EVA) methods have been used to model the relationship between flooding, watershed characteristics, and the weather, using regressions or hierarchical models to account for non-stationarity \citep{vsraj2016influence,dawdy2012regional,lima2016hierarchical}. However, purely statistical projections of extremal streamflow that do not consider physical variables which are expected to change under climate change (e.g., temperature and precipitation) are likely to be unreliable for long-term projections \citep{jain2001floods}. Precipitation has a large impact on groundwater flow and is therefore a major driver of extremal streamflow. Like streamflow, it exhibits non-stationarity \citep{Cheng2014,kunkel2020precipitation} which needs to be incorporated into any modeling that attempts to provide future projections of extremal streamflow. In this paper, our objective is to build a spatial model relating precipitation and streamflow and use climate model forecasts of future precipitation to understand flood risk under different climate change scenarios. 

While climate change is often described in terms of the mean, it will mostly be experienced through extremes. Data for extreme events are by definition sparse, and parametric models must therefore be carefully chosen based on extremal theory to estimate small probabilities. Standard measures of dependence such as correlation and spatial models such as Gaussian processes (GP) do not adequately model  extreme events; in order to properly account for spatial dependence while modeling rare event probabilities, we use spatial extreme value analysis (EVA). For modeling block maxima, e.g., the annual maximum of daily streamflow, the commonly used spatial EVA model is the max-stable process (MSP) \citep{de2006extreme,Smith-1990, Tawn, Schlather, Kabluchko-Schlather-deHaan, Buishand-etal, Wadsworth-Tawn-2012,Reich-Shaby}. MSPs are a natural asymptotic model for block maxima, but can also be applied to peaks over a threshold using a censored likelihood \cite[e.g.,][]{Huser-Davison-2014,Reich-Shaby-Cooley}. Exact inference for MSPs is challenging, and commonly used censored likelihood models for MSPs are also computationally intractable for all but a small number of spatial locations \citep{Schlather, Kabluchko-Schlather-deHaan, Wadsworth-Tawn-2012,wadsworth-2014a,wadsworth-2015a}. Further, MSPs enforce asymptotic dependence among spatial locations \citep{Huser-Wadsworth}, an unreasonable assumption for environmental data that often has weakening spatial dependence with increasing extremeness. Alternatives and extensions to MSPs include process mixture models \citep{Huser-Wadsworth, MajumderReichShaby2022,ZhangetalJASA2022} and max-infinitely divisible process (MIDP) models \citep{BoppMIDP2021JASA}, both of which can accommodate more flexible asymptotic regimes of tail dependence.

Climate-informed flood projections which consider non-stationarity is an ongoing area of research  \citep{delgado2014projecting,condon2015climate,franccois2019design,schlef2018general,schlef2021comparing,sankarasubramanian2003flood,zhang2015evaluation,bertola2019informed,Awasthietal2022}, but flood projections are not commonly studied as a spatial EVA problem. The intractability of common spatial EVA likelihoods pose computational challenges which make it difficult to fit realistic statistical models. For example, in a study of a large geographic region under a changing climate, it is unrealistic to assume stationarity in the degree of extremal dependence between nearby locations. Non-stationarity could appear due to dependence on variables which vary spatio-temporally, or due to physical considerations like topography. Recent work on incorporating non-stationarity in spatial EVA models include \cite{WADSWORTH2022100677}, which incorporates non-stationarity using the framework of \cite{sampsonGuttorp}. They deform the coordinate system into one where the process is stationary; this approach, however, does not use covariates. \cite{Huser2016} use covariates in the covariance structure of an MSP, extending the work of \cite{paciorek}. \cite{chevalier2021modeling} uses multidimensional scaling to capture regional variation in the asymptotic spatial dependence of an MSP, and \cite{ZhongetalAoAS2022} construct an MIDP which include covariates to capture spatio-temporal non-stationarities. Similarly, our work proposes a spatial EVA model that allows extremal dependence to vary over both space and time via climate covariates. While this model is flexible and intuitive, it is difficult to fit using standard computational methods. 

Many of the modeling and computational limitations of extreme value theory have been addressed using deep learning.  For example,  \citep{cannon2010flexible,vasiliades2015nonstationary,shrestha2017projecting,pasche2022neural,richards2022unifying} and the references therein use neural networks to obtain flexible regression frameworks relating covariates to extreme quantities.  Similar to the application in this paper, \cite{shrestha2017projecting} use neural networks to model the dependence of extreme streamflow and precipitation and temperature, and then use these relationships with climate models to project future extreme streamflow events.  Recently, \cite{wilson2022deepgpd} have used a convolutional neural network to regress spatial fields onto the parameters of an extreme value distribution.  Computational limitations due to intractable likelihoods associated with spatial extreme value processes have also been addressed using deep learning. \cite{lenzi2021neural,Sainsbury-Daleetal} replace maximum likelihood estimators with neural networks, while
\cite{MajumderReichShaby2022} develop synthetic likelihood functions by sampling from the spatial extreme value process with different parameter settings, and fitting these simulations with neural networks to learn an approximate likelihood function connecting the data with the model parameters.

In this work, we propose a non-stationary process mixture model (NPMM) for climate-informed estimation of extremal streamflow. We specify a statistical EVA model for annual streamflow maxima within the central US (CUS) region, and use downscaled and bias-corrected precipitation projections obtained from the Multivariate Adaptive Constructed Analogs (MACA) dataset \citep{MACA} as predictors. 
The NPMM addresses two important aspects of climate-informed EVA modeling. First, the process mixture model allows learning both the type and strength of asymptotic (in)dependence from the data by interpolating between a GP and an MSP.
Second, the NPMM introduces non-stationarity by allowing the asymptotic regime of the spatial process to vary spatio-temporally as a function of precipitation for sub-regions within the CUS. Climate models not only consider different distributions of climate variables between historical and future time periods, they also consider multiple future pathways where model outputs diverge considerably as we extend the time horizon. Covariates allow us to accommodate potential changes in the spatial or marginal behavior or both for extreme streamflow under future climate projections which deviate from historical patterns.
Inference for the NPMM is separated into density estimation and parameter estimation. The density estimation, used to approximate the intractable likelihood of the spatial process, is carried out using semi-parametric quantile regression (SPQR) \citep{xu-reich-2021-biometrics}. The quantile process has a basis function representation, whose weights are estimated using a feed-forward neural network. The NPMM provides a flexible framework for incorporating covariates into the spatial process as well as the marginal distributions at each location, and we use it to project extremal streamflow for 2006--2035 informed by climate model precipitation under two different climate pathways.

The rest of the paper is organized as follows. We introduce the streamflow and precipitation datasets for the CUS region in Section 2. Section 3 presents the NPMM and discusses density estimation, parameter estimation, and tail behavior for the model. Density estimation using SPQR for the CUS locations is carried out in Section 4, and we conduct a simulation study to see how errors in density estimation affect parameter estimates. The analysis of extremal streamflow as a function of precipitation is presented in Section 5, along with future projections of extremal streamflow based on downscaled and bias-corrected climate model precipitation data. Section 6 concludes. 
\section{Hydroclimatic data for the Central US}\label{s:data}
\subsection{Observed streamflow data}
\begin{figure}
    \centering
    \includegraphics[scale=0.6]{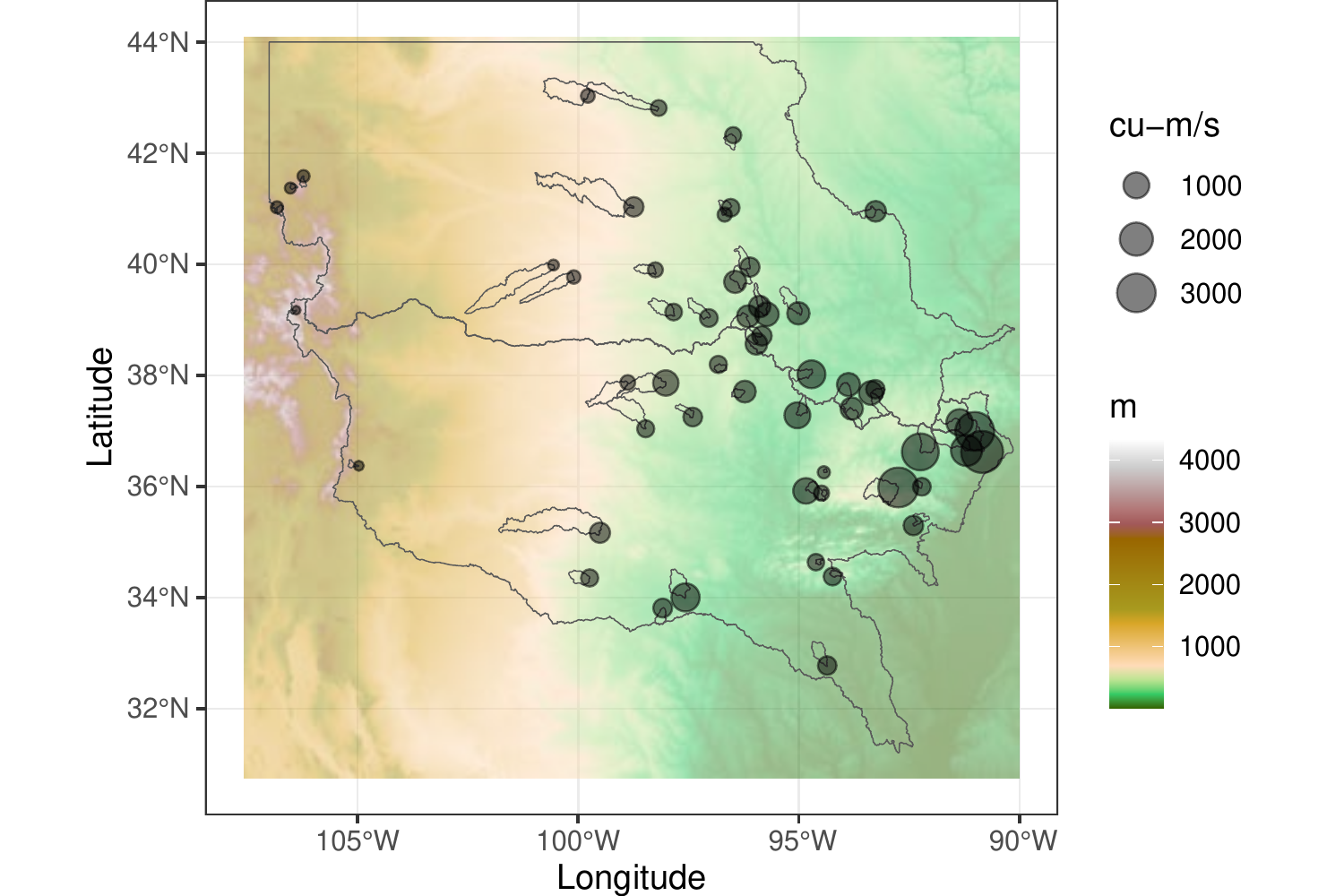}
    \caption{\small \textbf{HCDN sites in HUC-02 regions 10L and 11}: Locations and 0.99 quantiles of annual streamflow maxima (in $m^3/s$) at 55 HCDN stations overlaid on an elevation map (in $m$) of the central United States region bounded by $[-107, -90] \times [30, 44]$. The two large polygons within the figure correspond to regions 10L (top) and 11 (bottom), and the smaller polygons correspond to the HCDN basins that each station measures streamflow for.}
        \label{fig:99pctl}
\end{figure}

The USGS Hydro-Climatic Data Network (HCDN) \citep{lins2012usgs} is a dataset of streamflow records within the United States and its Territories. The HCDN consists of locations that are minimally impacted by anthropogenic activity, making it suitable to study the effects of changing climate on streamflow. The HCDN has been used to study the effect of climatic variables on streamflow \citep{Sankaretal2001,OhSankar-2012} and to study the change in extremal streamflow over time \citep{MajumderReichShaby2022}. For studying water resources, the USGS divides the US into groups of nested Hydrologic Units, identified by Hydrologic Unit Codes (HUCs). The first level of classification divides the US into 21 regions, referred to as HUC-02 regions. Our study focuses on two specific HUC-02 regions for which we have a complete data record between 1972--2021; the lower half of Region 10, denoted as 10L, and Region 11. Together, they span a region in the Central US (CUS) that consist of 55 gauges spread across South Dakota, Nebraska, Colorado, Wyoming, Kansas, Iowa, Missouri, Arkansas, Oklahoma, New Mexico and Texas. Figure \ref{fig:99pctl} plots the sample 0.99 quantile of annual streamflow maxima (measured in $m^3/s$) over the last 50 years at each station. There is spatial variation in these data, with extremal streamflow increasing from west to the east.

\subsection{Observed precipitation data} 
The CUS is characterized by severe convective storms \citep{Risseretal2019,ZHANG2022100499}, and precipitation trends that could potentially influence flooding \citep{kunkel2020precipitation}. \citep{condon2015climate} have used monthly average precipitation as a model predictor to project future floods, while \citep{Awasthietal2022} have used monthly total precipitation as predictors to project flood frequencies under near-term climate change. We refer the reader to \citep{Awasthietal2022} for further references regarding the use of precipitation and temperature as predictors of extremal streamflow. In this study, we use seasonal and annual precipitation means as predictors of annual extremal streamflow. Monthly precipitation data is sourced from the NOAA Monthly US Climate Gridded Dataset (NClimGrid) \citep{NClimGrid}, which is based on the Global Historical Climatology Network (GHCN) dataset. NClimGrid data is available on a 5km $\times$ 5km grid, and for each of the 55 HCDN stations we use monthly precipitation for all NClimGrid cells for the corresponding basins as outlined in Figure \ref{fig:99pctl}.

\begin{figure}
    \centering
    \begin{subfigure}{\linewidth}
    \includegraphics[scale=0.75]{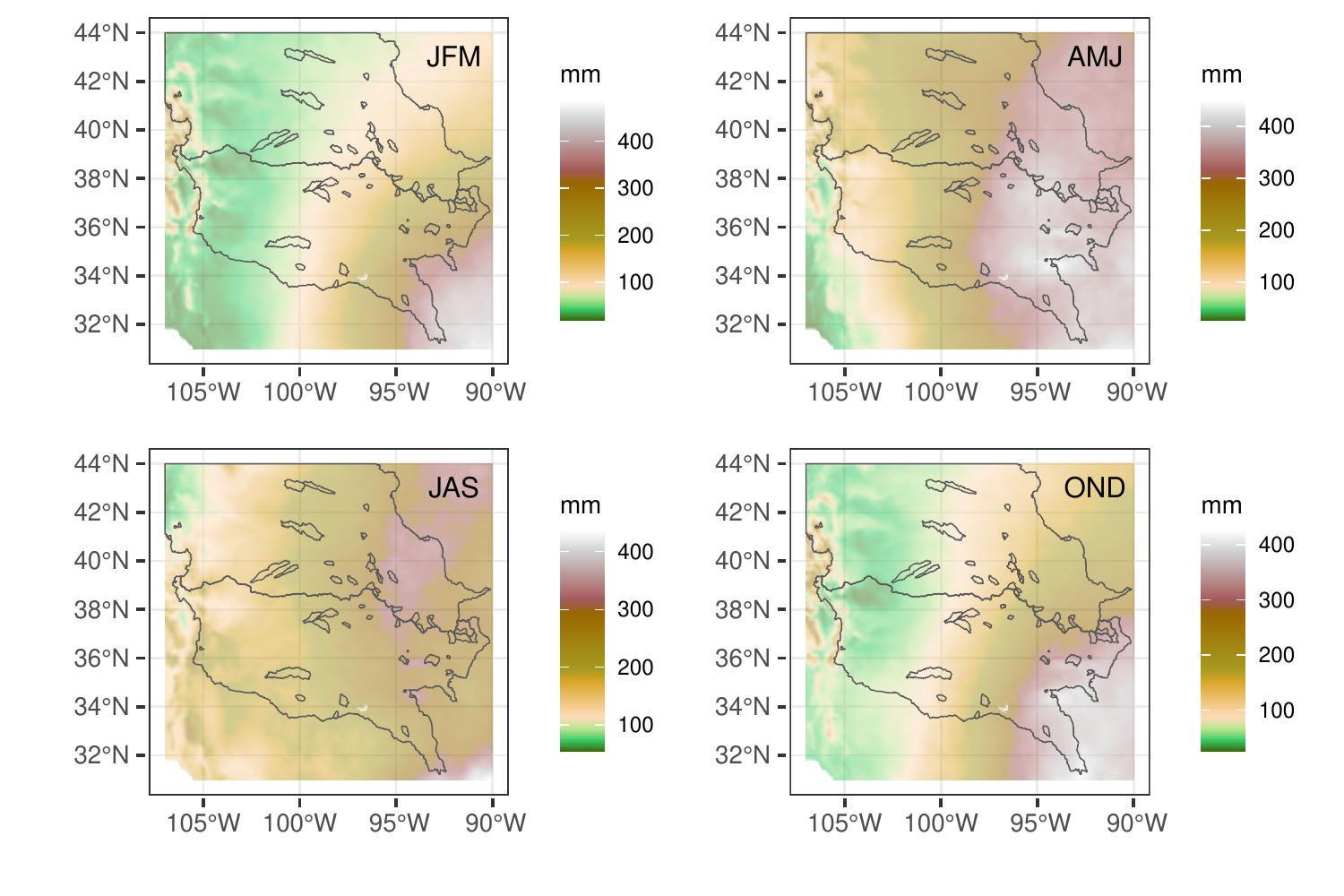}
\caption{\small Mean of seasonal precipitation over the CUS.}
    \label{fig:prcp_seasonal_grid}
    \end{subfigure}
    \hfill
    \begin{subfigure}{\linewidth}
    \includegraphics[scale=0.75]{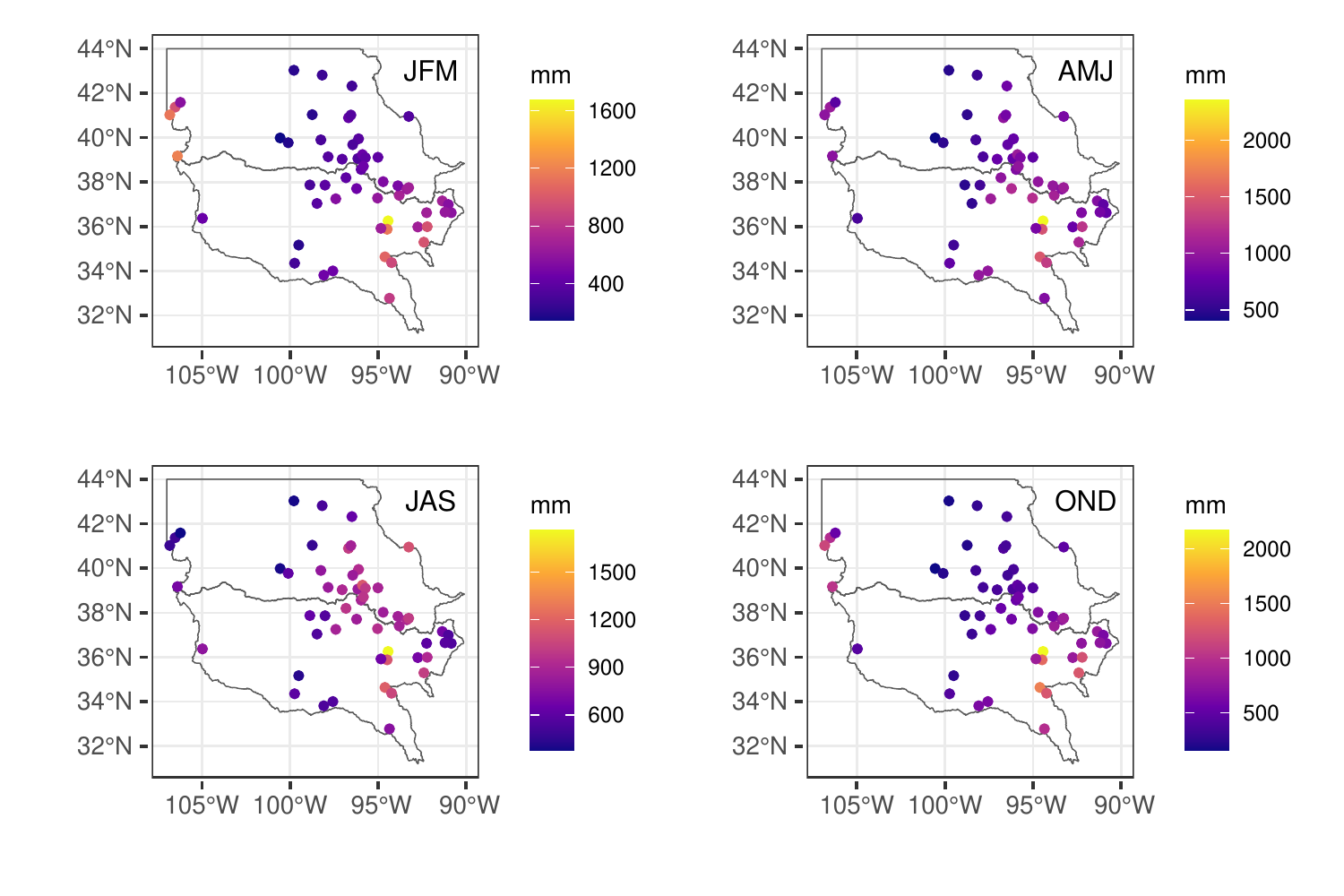}
\caption{\small 0.99 quantiles of seasonal precipitation associated with each HCDN site.}
    \label{fig:prcp_seasonal}
    \end{subfigure}
    \caption{\small \textbf{Seasonal distribution of NClimGrid precipitation for 1972--2021:} Seasons are specified on the top right of each panel and defined as winter (JFM), spring (AMJ), summer (JAS), and autumn (OND).}
    \label{fig:nclimgrid}
\end{figure}

\begin{figure}
    \centering
     \includegraphics[scale=0.7]{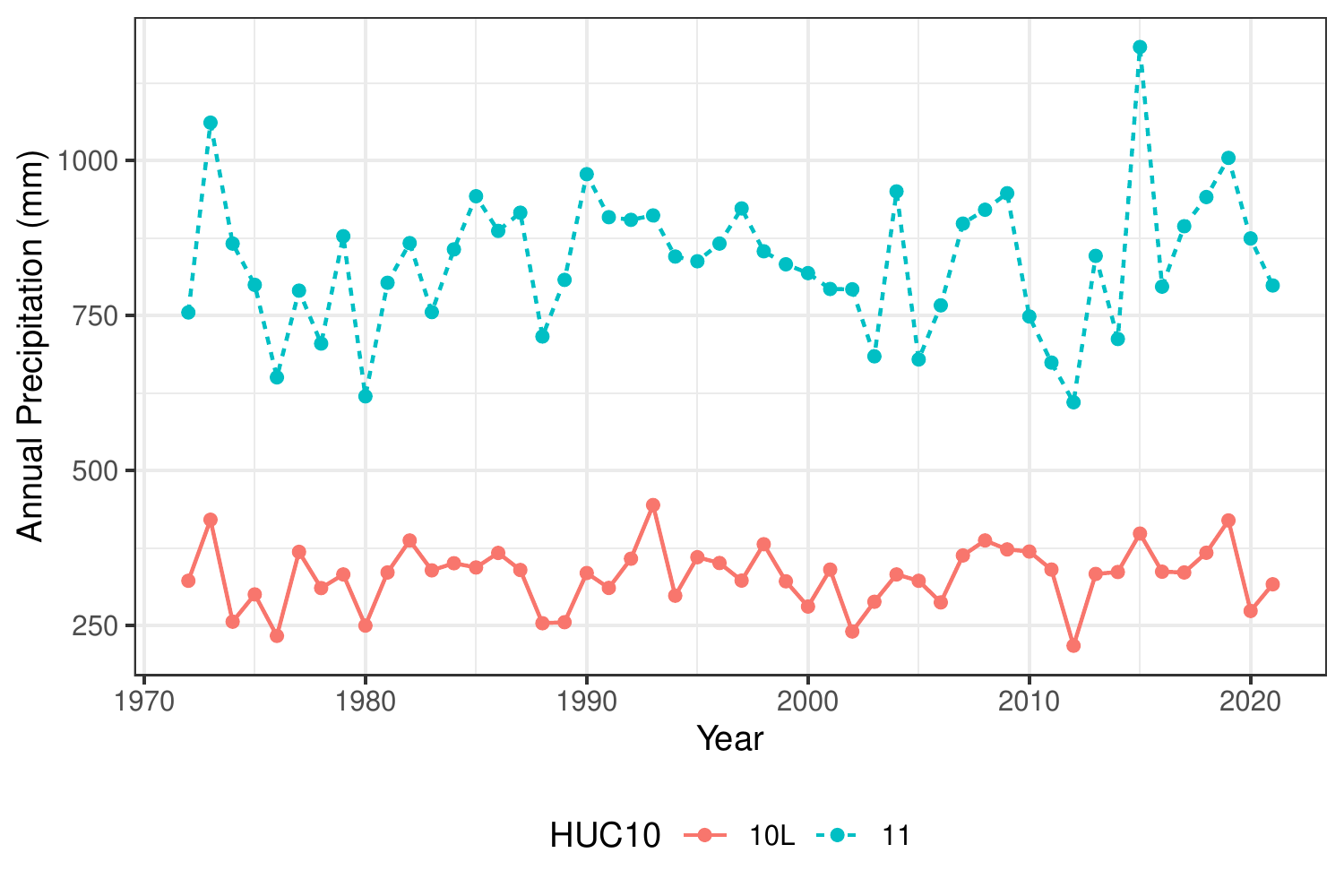}
        \caption{\small Time series of annual NClimGrid precipitation (in $mm$) from 1972--2021 for the 2 HUC-02 regions of the CUS. Values represent an average over all grid cells within the corresponding region.}
    \label{fig:basinprcp}
\end{figure}

The NClimGrid data are treated as covariates to estimate both the marginal parameters at each site as well as dependence parameters for the underlying spatial process. For our response variable $Y_t(\bs)$, the extremal streamflow for year $t$ and location $\bs$, we consider 
the corresponding seasonal precipitations  as covariates. Following \citep{Awasthietal2022}, the seasons correspond to winter (JFM), spring (AMJ), summer (JAS), and autumn (OND), where JFM denotes the months of January-February-March, and so on. Figure \ref{fig:prcp_seasonal_grid} plots the mean seasonal precipitation across the 2 HUC-02 regions. Not only is there spatial variability within a season, we also see heterogeneity across seasons. The highest values are observed in the southeast, and lower values seen along the west. We also note that the spring season has the highest precipitation. Figure~\ref{fig:prcp_seasonal} plots the 0.99 quantile of the seasonal precipitation associated with the HCDN sites for each season. Additionally, we define the covariates $Z_{1t}$ and $Z_{2t}$ as the annual precipitation within HUC-02 Regions 10L and 11, respectively, which is computed as the total precipitation for all NClimGrid points for the corresponding region. Figure \ref{fig:basinprcp} plots a time series of annual precipitation for the 2 HUC-02 regions from 1972--2021. We note that Region 11, which is located in the southern part of the CUS, has higher precipitation than Region 10L.

\subsection{Global Climate Model output of future precipitation}\label{s:maca}

\begin{figure}
    \centering
     \includegraphics[scale=1.2]{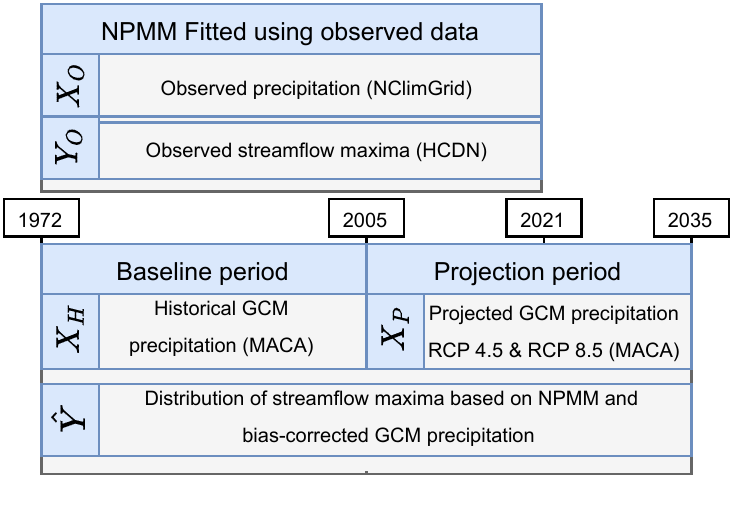}
        \caption{\small Datasets used in the study, with periods of availability and usage details.}
    \label{fig:calibration}
\end{figure}

While Global Climate Models (GCMs) do not produce streamflow estimates, they provide precipitation variables which we use to predict extremal streamflow.  The Multivariate Adaptive Constructed Analogs (MACA\footnote{\url{https://www.climatologylab.org/maca.html}}) dataset \cite{MACA} is a statistical downscaling method for GCMs. MACA downscales the model output from 20 GCMs of the Coupled Model Inter-Comparison Project~5 (CMIP5) \cite{CMIP5} for historical GCM forcings (1950--2005) as well as future Representative Concentration Pathways (RCPs) RCP 4.5 and RCP 8.5 scenarios (2006--2100) from the native coarse resolution of the GCMs to a higher spatial resolution of 4km. RCP 4.5 assumes that total anthropogenic CO$_2$ will peak around 2040, and decline till 2080, whereas RCP 8.5 assumes that CO$_2$ concentrations continue to rise until the end of the century. MACA provides monthly precipitation (\texttt{pr}) as one of its outputs; we obtain both the historical runs for 1972--2005 for calibrating it to NClimGrid output, and use it to estimate extremal streamflow for the CUS from 2006--2035. The quality of the GCM model projections can vary according to variable, climate pathway, geographic region, and time horizon, and \cite[page 9]{Joyce_2020} provides criteria for selecting climate models. We choose 6 models for each RCP scenario following on the model rankings provided by \cite{Joyce_2020}; the chosen models are the top three ranked projections in terms of precipitation change (dry, wet) at mid-century (2041--2070) under the 2 scenarios (RCP 4.5, RCP 8.5) at the coterminous US scale. While our study focuses on projections up until 2035, our choice of models ensure that these results can be extended for longer durations, and account for model and scenario uncertainty. The models chosen for RCP 4.5 are IPSL-CM5A-MR, bcc-csm1-1-m, IPSL-CM5A-LR, CSIRO-Mk3-6-0, CNRM-CM5, and MRI-CGCM3; models chosen for RCP 8.5 are IPSL-CM5A-MR, HadGEM2-ES, inmcm4, CNRM-CM5, MRI-CGCM3, and CSIRO-Mk3-6-0. We refer readers to \cite{Joyce_2020} for further comparisons of all 20 models. Figure \ref{fig:calibration} contains a schematic of the observational and climate model datasets used in this study, as well as the historical and projection time periods.

\begin{figure}
    \centering
     \includegraphics[scale=0.8]{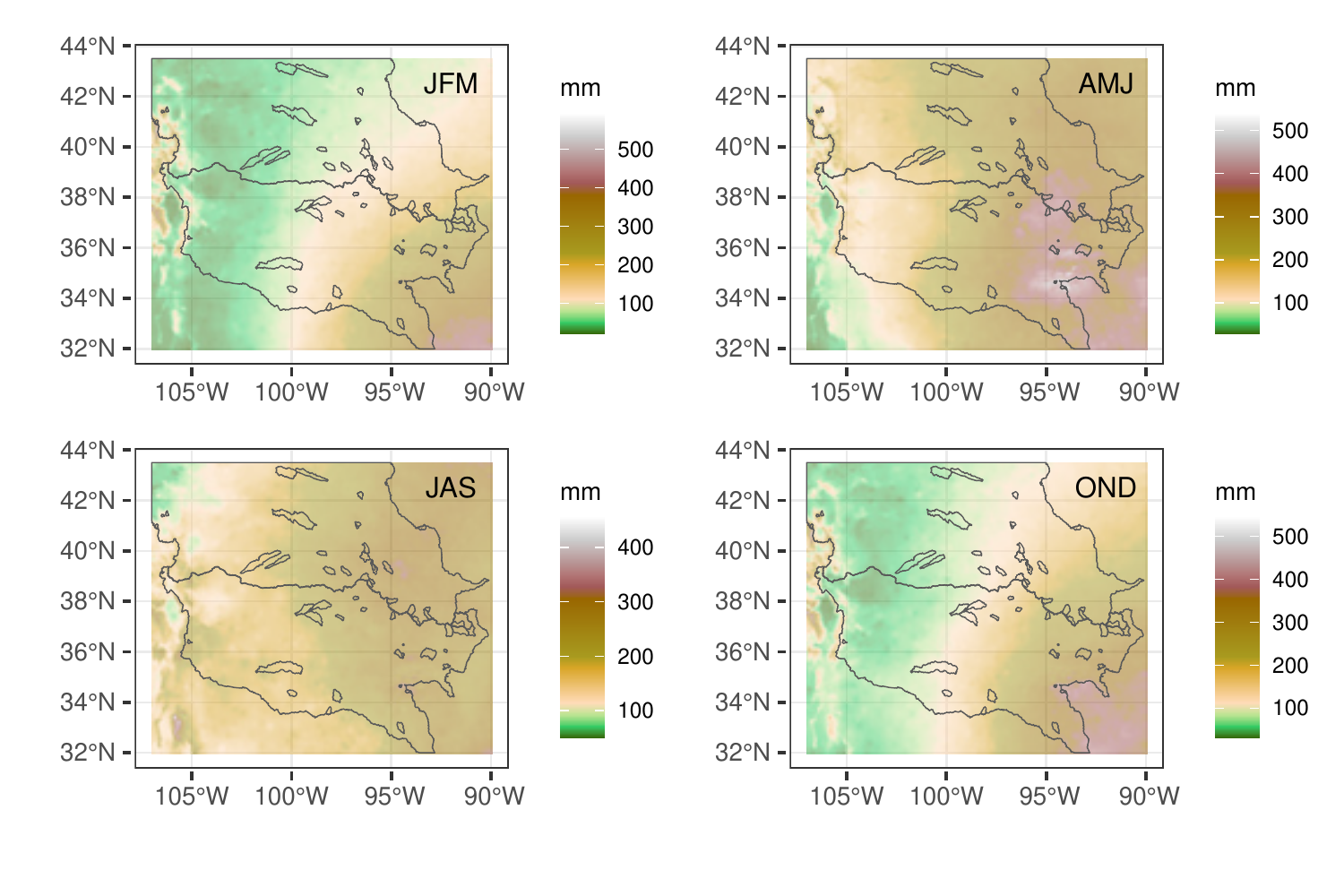}
        \caption{\small Mean seasonal precipitation (in $mm$) over the CUS based on the CNRM-CM5 model for the GCM historical period of 1972--2005. Seasons are specified on the top right of each panel and defined as winter (JFM), spring (AMJ), summer (JAS), and autumn (OND).}
    \label{fig:CNRM_CM5_rast}
\end{figure}

The GCM data do not have temporal correspondence; GCM output for the year 2005 is not a representation of the weather in 2005. Rather, GCM data for the historical and future periods are designed to approximate the distribution of the observed or forecast data for similar time periods. The lack of temporal correspondence makes it inappropriate to regress observed streamflow onto modeled precipitation to estimate the relationship between these variables. However, given a model fit using temporally-correspondent observed precipitation and streamflow, estimates generated using bias-corrected GCM data as covariates can be used to compare the changes in the distribution across different time periods. Figure \ref{fig:CNRM_CM5_rast} plots mean seasonal precipitation over the CUS based on the CNRN-CM5 model for the GCM historical period of 1972--2005. This is one of the models projecting a wetter future, and the historical precipitation from this model is higher than the observed NClimGrid data in Figure \ref{fig:prcp_seasonal_grid}. The spatial patterns are broadly similar between the two datasets, and the GCM output needs to be calibrated to the observational data before it can be used as a covariate to model extremal streamflow.

The GCM output is calibrated to remove bias compared to the observed precipitation at each location.  The GCM log-precipitation outputs during the historical period are calibrated to have the same sample mean and variance as the observed precipitation for the same time period.  This log-linear transformation is estimated and applied separately for each HCDN station and each GCM forcing, and applied to the GCM precipitation projections as well. Precipitation for the 2 HCDN regions are also similarly calibrated. It is recommended that averages of the weather over at least 30 years be used to assess the climate. Hence, we consider a historical (baseline) period of 1972--2005 and a future projection period of 2006--2035, and study changes in the extremal quantiles of the distribution of predicted streamflow maxima for these two time periods.

\section{Non-stationary Process Mixture Models for Spatial Extremes}
\label{s:model}

\subsection{The NPMM for block maxima}\label{s:npm}
Let $Y_t(\bs)$ be the extreme observation at time $t$ and spatial location $\bs$, for $t \in \{1,\ldots, T\}$ and $\bs \in \{\bs_1,\ldots, \bs_n\}$. The observations $Y_t(\bs)$ are defined as block maxima, and are thus assumed to arise from a generalized extreme value (GEV) distribution with location $\mu_t(\bs)$, scale $\sigma_t(\bs)$, and shape $\xi_t(\bs)$:
$$
  Y_t(\bs) \sim \mbox{GEV}\{\mu_t(\bs),\sigma_t(\bs),\xi_t(\bs)\},
$$
whose cumulative distribution function (CDF) $F_{t, \bf s}(y) := \mathbb{P}[Y_t(\bs)<y]$ is
\begin{equation}
    \mathbb{P}\bigl[Y_t(\bs)<y\bigr] = \exp \biggl\{-\left[1+\xi_t(\bf s)\left(\frac{y-\mu_t(\bf s)}{\sigma_t(\bf s)}\right)\right]^{-1/\xi_t(\bf s)}\biggr\}.
\end{equation}
The CDF is defined over the set $\bigr\{y:1+\xi_t(\bs)(y-\mu_t(\bs))/\sigma_t(\bs)>0\bigr\}$. 

Denote $Z_{jt},$ for $ j=1,2$ and $t=1,\ldots,50,$ as the annual precipitation for the two HUC-02 regions (10L and 11) defined in Section \ref{s:data}. We define $X_{1t}(\bs)$ as the annual precipitation for the HUC-02 region that location $\bs$ belongs to, i.e.,
$$X_{1t}(\bs) = \mathbb{I}\{\bs \in \mbox{Region 10L}\}Z_{1t} + \mathbb{I}\{\bs \in \mbox{Region 11}\}Z_{2t},$$
where $\mathbb{I(\cdot)}$ is the indicator function. Further, denote $X_{it}(\bs), i=2,\ldots,5$ and $t=1,\ldots,50$ as the seasonal precipitation for site $\bs$ at time $t$ for the four seasons as defined in Section \ref{s:data}. We assume the GEV location parameters vary spatially and are dependent on precipitation, while the scale and shape parameters also vary spatially, i.e.,
\begin{align}\label{e:marginal}
    &\mu_t(\bs) = \mu_0(\bs) + \sum_{i=1}^5\mu_i(\bs)X_{it}(\bs), 
    &\sigma_t(\bs) &= \sigma(\bs).
    &\xi_t(\bs) &= \xi(\bs).
\end{align}
The CDF transformed variables
$U_t(\bs) := F_{t,\bf s}\bigl(Y_t(\bs)\bigr)$ share common uniform marginal distributions but are spatially correlated; this transformation separates residual spatial dependence in $U_t(\bs)$ from the spatial dependence induced by spatial variation in the GEV parameters, which can be modeled using GP priors over $\bs$. 

A spatial dependence model on $U_t(\bs)$ is obtained via the transformation $U_t(\bs) = G_{t,\bf s}\bigl(V_t(\bs)\bigr)$, such that
\begin{equation} 
  \label{eq:copula_pareto-02}
  V_t(\bs) = \delta_t(\bs) g_R\bigl(R_t(\bs)\bigr) + (1-\delta_t(\bs))g_W\bigl(W_t(\bs)\bigr),
\end{equation}
where $R_t(\bs)$ is a max-stable process (MSP), $W_t(\bs)$ is a Gaussian process (GP), and $g_R$ and $g_W$ are transformations to ensure that $g_R\bigl(R_t(\bs)\bigr)$ and $g_W\bigl(W_t(\bs)\bigr)$ both follow the standard exponential distribution. Without loss of generality, we assume that $R_t(\bs)$ has a marginal GEV$(1,1,1)$ distribution and $W_t(\bs)$ has a  marginal N$(0,1)$ distribution; the corresponding transformations are $g_R(r) =  -\log(1-\exp(-1/r))$ and $g_W(w) = -\log(1-\Phi(w))$ for the standard normal CDF $\Phi(w)$. By construction, $V_t(\bs)$ follows a two-parameter hypoexponential distribution marginally, with CDF 
\begin{equation}
    \label{eq:hypo}
    G_{t,\bf s}(v) = 1 - \frac{1-\delta_t(\bs)}{1-2\delta_t(\bs)}e^{-\frac{1}{(1-\delta_t(\bf s))}v} + \frac{\delta_t(\bs)}{1-2\delta_t(\bs)}e^{-\frac{1}{\delta_t(\bf s)} v}.
\end{equation} 
The parameters $\delta_t(\bs) \in[0,1]$ are weight parameters that control the relative contribution of the two spatial processes at every site and time point.

The spatial dependence model in \eqref{eq:copula_pareto-02} was originally introduced in \citep{MajumderReichShaby2022} where it assumed a constant value of $\delta_t(\bs) = \delta$. In practice, however, it is reasonable to partition the sites into $L$ regions such that sites within each partition share a common value of $\delta_t(\bs)$ at any given time point $t$, with different partitions having potentially different values of $\delta_t(\bs)$. Locations can be assigned to partitions based on underlying geophysical characteristics of the data, or clustered according to an appropriate distance metric. For streamflow data, the two HUC-02 regions (10L and 11) are considered partitions of the CUS. Thus $L = 2$ for our study, and we denote $\delta_{1t}$ and $\delta_{2t}$ as the weight parameters for these 2 partitions, i.e.,
$$\delta_t(\bs) = \mathbb{I}\{\bs \in \mbox{Region 10L}\}\delta_{1t} + \mathbb{I}\{\bs \in \mbox{Region 11}\}\delta_{2t}.$$
As with the marginal parameters, we assume $\delta_{1t}$ and $\delta_{2t}$ depend on partition-specific covariates:
\begin{align}\label{e:joint}
    g^{-1}(\delta_{it}) = \beta_{i0} + \beta_{i1} Z_{it}, i = 1,2,
\end{align}
where $g(\cdot)$ is an appropriate link function, and $Z_{it}$ are the annual precipitation for the two HUC-02 regions as defined in Section \ref{s:data}. The variable $\delta_{it}$ depends on time through the covariate $Z_{it}$. Mixing the asymptotically dependent MSP with the asymptotically independent GP provides a rich model for spatial dependence, while the covariates help capture changes in the spatio-temporal dependence.

We model the correlation of the GP $W_t(\bs)$ using the isotropic powered-exponential correlation function $\mbox{Cor}\bigl(W_t(\bs_1),W_t(\bs_2)\bigr) = \exp\{-(h/\rho_W)^{\alpha_W}\}$ with distance $h=||\bs_1-\bs_2||$, smoothness $\alpha_W\in(0,2)$, and range $\rho_W>0$. The MSP $R_t(\bs)$ is assumed to have isotropic Brown-Resnick spatial dependence defined by the variogram $\gamma(h) = (h/\rho_R)^{\alpha_R}$ for smoothness $\alpha_R\in(0,2)$ and range $\rho_R>0$. We also incorporate a nugget into the process mixture. We denote the proportion of the variance explained by the spatial process by $r$, and construct $W_t(\bs)$ and $R_t(\bs)$ as:
\begin{align*}
    \mbox{Cor}\bigl(W_t(\bs_1),W_t(\bs_2)\bigr) &= r\cdot\exp\{-(h/\rho_W)^{\alpha_W}\}\\
    R_t(\bs) &= \max\{r\cdot R_{1t}(\bs),(1-r)\cdot R_{2t}(\bs)\},
\end{align*}
where $R_{1t}(\bs)$ is an MSP, and $R_{2t}(\bs) \iid \mbox{GEV}(1,1,1)$ distributed independently of $R_{1t}(\bs)$. 

We refer to this model as a non-stationary process mixture model (NPMM), with marginal parameters $\btheta_1 = \{\mu_0(\bs_i),...,\mu_5(\bs_i),\sigma(\bs_i),\xi(\bs_i);i = 1:n\}$ and spatial dependence parameters $\btheta_2=\{\beta_{10},\beta_{11},\beta_{20},\beta_{21},\rho_R,\alpha_R,\rho_W,\alpha_W,r\}$. Alternative spatial dependence structures are viable under the NPMM; in general, most spatial processes are compatible with the methodology presented in this work. For the purposes of this particular problem, we choose a relatively smooth spatial process, and aim to capture additional complexity using spatio-temporally varying coefficients (STVC) models \cite{SVC,MajumderReichShaby2022} on the components of $\btheta_1$.
 
\subsection{Asymptotic joint tail behavior for the NPMM}\label{s:tail_behavior}

Extremal spatial dependence of the process at sites $\bs_1$ and $\bs_2$ is often measured using the conditional exceedance probability,
\begin{equation}\label{e:chi_defn}
    \chi_u(\bs_1,\bs_2) := \mathbb{P}\{U(\bs_1) > u | U(\bs_2) > u\},
\end{equation}
where $u\in(0,1)$ is a threshold. The random variables $U(\bs_1)$ and $U(\bs_2)$ are defined as asymptotically dependent if the limit
\begin{equation}
    \chi(\bs_1,\bs_2) = \lim_{u\rightarrow 1}\chi_u(\bs_1,\bs_2)
\end{equation}
is positive, and independent if $\chi(\bs_1,\bs_2) = 0$.


\begin{figure}
    \centering
    \begin{subfigure}[b]{0.49\linewidth}
    \includegraphics[width=\linewidth]{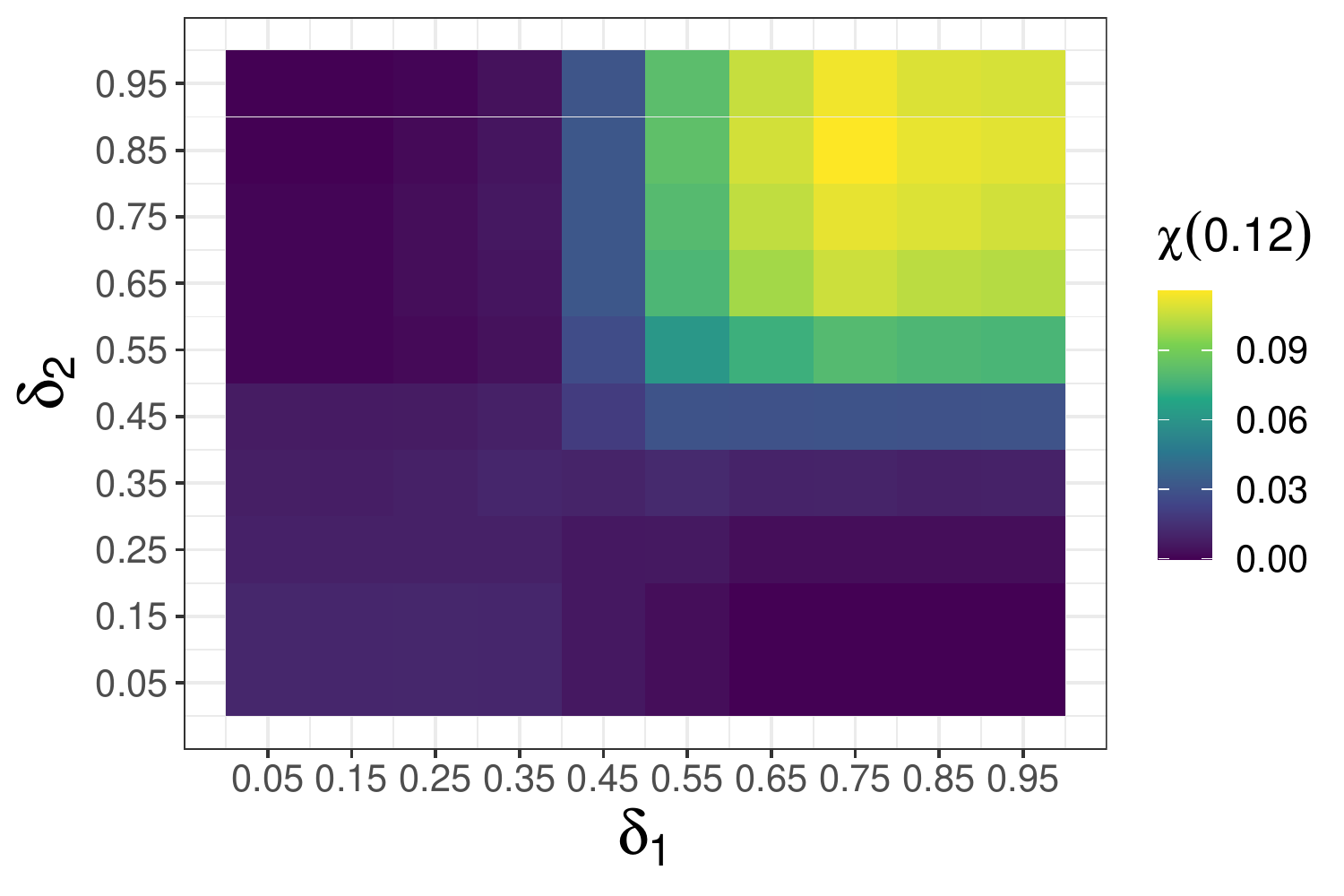}
\caption{\small $\chi_u(0.12)$ for different combinations of $\delta_1$ and $\delta_2$ with threshold $u=0.9999$.}
    \label{fig:chi_sq}
    \end{subfigure}
    \hfill
    \begin{subfigure}[b]{0.49\linewidth}
    \includegraphics[width=\linewidth]{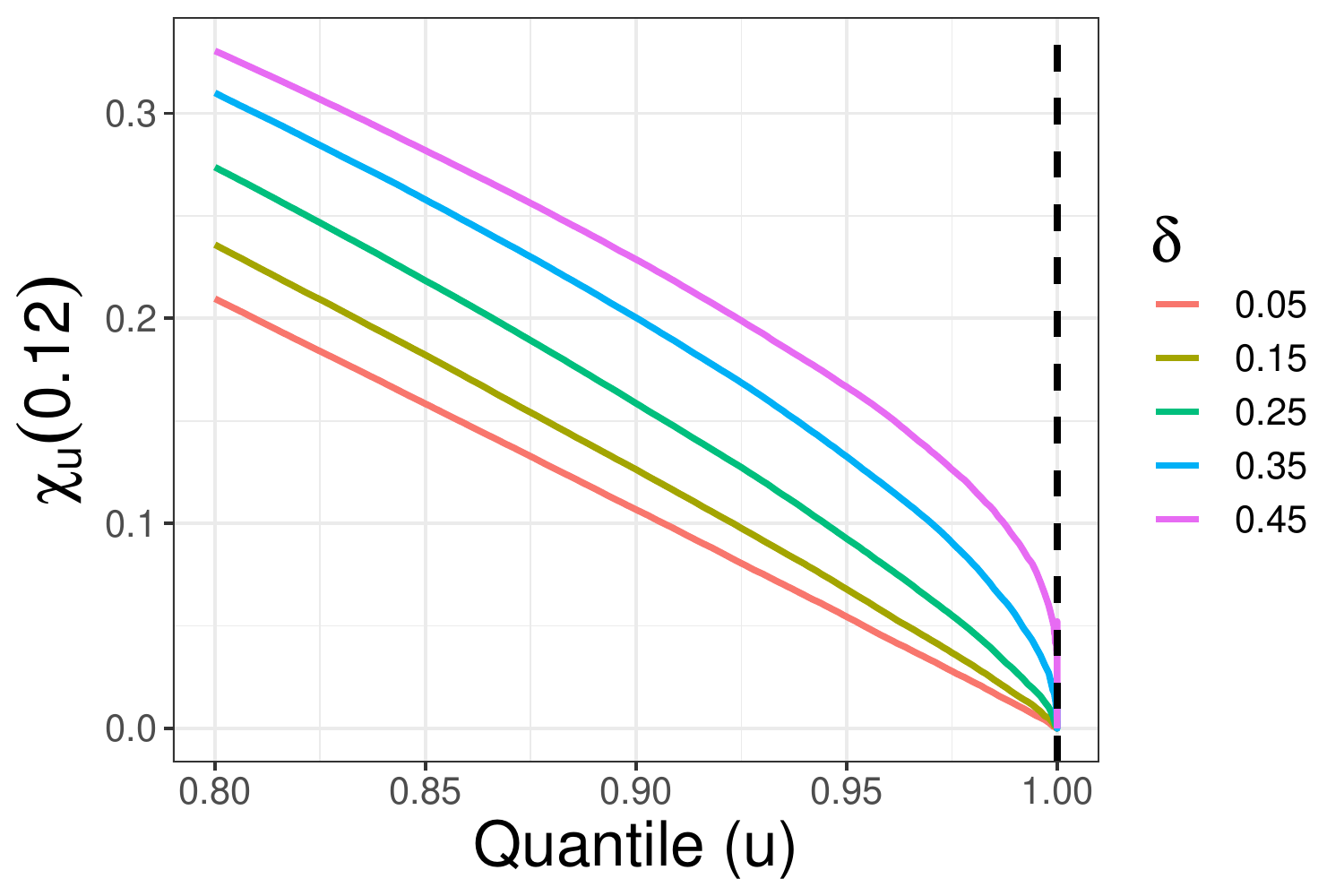}
\caption{\small $\chi_u(0.12)$ for different values of $\delta$ with $\delta_1=\delta$ and $\delta_2=1-\delta$.}
    \label{fig:chi_empirical}
    \end{subfigure}
    \caption{\small Empirical $\chi_u(h)$ where $h=0.12$ for the process mixture model as a function of $\delta_1$ and $\delta_2$ for sites corresponding to the HCDN stations in the CUS.}
    \label{fig:empericalchi}
\end{figure}

To examine the model in a simpler case, we assume $\delta_{1t}$ to be the same for $t = 1,\ldots, T$, and define $\delta_i := \delta_{it}, i=1,2$. We numerically approximate $\chi_{u}(\bs_1,\bs_2)$ for various values of $u$, $\delta_1$ and $\delta_2$. We scale our region of interest and all 55 sites within it to fall within the unit square, and consider the extremal spatial dependence between a hypothetical pair of sites at a distance of $h = 0.12$ from each other. The value for $h$ is chosen as the solution to:
\begin{align*}
    h = \max_{i = 1:55} ||\bs_i - \bs_{i^*}||,
\end{align*}
where $a:b$ is used as shorthand notation for $a, a+1,\ldots,b-1,b$, and $\bs_{i^*}$ is the site closest to $\bs_i$. In its original scale, this is equivalent to HCDN stations 218 km apart. Figure \ref{fig:chi_sq} plots the behavior of $\chi_u(\bs_1,\bs_2)$ for different $(\delta_1,\delta_2)$ pairs. Assuming an isotropic model, $\chi_u(\bs_1,\bs_2)$ is a function only of the distance $h = ||\bs_1-\bs_2||$, and so we use the notation $\chi_u(h):= \chi_u(\bs_1,\bs_2)$. While $\chi_u(h)$ depends on $(\delta_1, \delta_2)$ in our work, we suppress the dependence for notational convenience and instead use $\chi_u(h)$ in the remainder of the text. As in \citep{Huser-Wadsworth}, we set the GP to have a correlation of 0.40, which is equivalent to fixing $\rho_W = 0.134$ and $\rho_R = 0.19\rho_W$ (see Section \ref{s:numericalstudies} for a discussion on the choice of $\rho_W$ and $\rho_R$), and computed the conditional exceedance probability for $u=0.9999$. When $\delta_1=\delta_2 = \delta$, \citep{MajumderReichShaby2022} have shown using empirical studies that $\chi_u(h)\to 0$ if $\delta<0.5$ and $\chi_u(h)>0$ for $\delta>0.5$. An analytical result consistent with this finding was also derived for the case of a shared extremal process, i.e., for $R(\bs_1) = R(\bs_2) = R$, at which point we recovered the similar result from \citep{Huser-Wadsworth}. From Figure \ref{fig:chi_sq}, we can also see that $\chi_u(h)\to 0$ when both $\delta_1, \delta_2<0.5$. To understand the tail behavior of the process when $\delta_1$ is high and $\delta_2$ is low (and vice-versa), we consider the case where $\delta_1=\delta, \delta_2 = 1-\delta$, for $\delta\in (0,1)$. We find that $\chi(h)\to 0$ in this situation for all values of $\delta$; this is verified empirically in Figure \ref{fig:chi_empirical} where $\chi_u(h)\to 0$ for different values of $\delta_1 \mbox{ and } \delta_2$. It also corresponds to the diagonal in Figure \ref{fig:chi_sq} which is shown to go to 0. This is intuitively reasonable; $R(\bs)$ and $W(\bs)$ are independent, and thus asymptotic dependence is only achieved if both sites have large delta and thus both sites allow substantial contribution for the asymptotically dependent process $R(\bs)$. An analytical derivation of this result for the case of a shared extremal process is provided in \ref{s:chi_derivation}.

\subsection{Density regression using Deep Learning for the NPMM}

Assume the process is observed at $n$ sites $\bs_1,...,\bs_n$.  We partition the parameters into those that affect the marginal distributions in (\ref{e:marginal}), denoted $\btheta_1$, and those that affect the spatial dependence, denoted $\btheta_2$. Denoting $Y(\bs_i)\equiv Y_{i}$ and $U_{i}:=F(Y_{i};\btheta_1)$, we can express the joint distribution for all the observations using a change of variables, as:
\begin{equation}\label{e:changeofvariables}
  f_y(y_{1},...,y_{n};\btheta_1,\btheta_2) = f_u(u_{1},...,u_{n};\btheta_2)\prod_{i=1}^n\left|\frac{dF(y_{i};\btheta_1)}{dy_{i}}\right|.  
\end{equation}

Model fitting for the NPMM is challenging due to the way the spatial dependence is specified; the joint distribution of the MSP $R(\bs)$ is available only for a moderate number of locations, and working with the term $f_u(u_{1},...,u_{n};\btheta_2)$ in \eqref{e:changeofvariables} analytically is not viable. As in \citep{MajumderReichShaby2022}, the change of variables in \eqref{e:changeofvariables} sets the process mixture component up for density estimation. The density estimation is carried out on a surrogate likelihood based on a Vecchia decomposition \citep{vecchia1988estimation,stein2004approximating,datta2016hierarchical,katzfuss2021general} of the joint distribution $f_u(u_{1},...,u_{n};\btheta_2)$,
\begin{equation}\label{e:vecchia}
    f_u(u_{1},...,u_{n};\btheta_2) = \prod_{i=1}^n f_i(u_{i}|\btheta_2,u_{1}, ...,u_{i-1})
    \approx
    \prod_{i=1}^n f_i(u_{i}|\btheta_2,u_{(i)}),
\end{equation}
for $u_{(i)} = \{u_j; j\in \calN_i\}$ and $\mathcal{N}_i \subseteq \{1,\ldots,i-1\}$. The set of locations $\bs_{(i)}$ are analogously defined as $\bs_{(i)} = \{\bs_j; j\in \calN_i\}$ and is referred to as the Vecchia neighboring set. The approximation therefore entails truncating the dependence that $u_i$ has on all its previous $i-1$ ordered sites to instead consider dependence on only up to $m$ sites, i.e., $|\calN_i| \leq m$. The first term of the approximation is the marginal density $f_1(u_1|\btheta_2)$.

The univariate conditional distribution terms on the right hand side of \eqref{e:vecchia} do not have closed-form expressions.
Density regression is carried out for each of the $n-1$ terms separately using the semi-parametric quantile regression (SPQR) model introduced in \citep{xu-reich-2021-biometrics}:
\begin{equation}\label{e:1}
    f_i(u_{i}|\bx_{i},\mathcal{W})= \sum_{k=1}^{K}\pi_{ik}(\bx_{i},\mathcal{W}_i)B_{k}(u_{i}),
\end{equation}
for $i = 2:n$, where $\pi_{ik}(\bx_i,\mathcal{W}_i)\ge 0$ are probability weights with $\sum_{k=1}^K\pi_{ik}(\bx_i)=1$ that depend on the parameters $\mathcal{W}_i$, and $B_k(u_i)\ge0$ are M-spline basis functions that, by definition, satisfy $\int B_k(u)du=1$ for all $k$. The density regression model in \eqref{e:1} treats $u_{(i)}$ and $\btheta_2$ as features (covariates), denoted as $\bx_i$, with $u_i$ being the corresponding response variable.

By increasing the number of basis functions $K$ and appropriately selecting the weights $\pi_{ik}(\bx_i)$, the mixture distribution in \eqref{e:1} can approximate any continuous density function \citep[e.g.,][]{chui1980,abrahamowicz1992} which makes it suitable for our application.
The weights are modeled using a neural network (NN) with $H$ hidden layers and a multinomial logistic (softmax) activation function on its output layer, i.e.,
\begin{equation}\label{e:ffnn}
 \pi_{ik}(\bx_i,\mathcal{W}_i) = f^{NN}_i(\bx_i,\mathcal{W}_i),\mbox{ for }i = 2,\ldots,n.
\end{equation}

Instead of using observational data, the weights are learned from training data generated from the process mixture model with parameters $\btheta_2 \sim p^*$, which can then be used to obtain realizations from the process over sites $\bs_i$ and $\bs_{(i)}$ from the model conditioned on $\btheta_2$. Specifically, we generate data at the observed spatial site with the same Vecchia neighbor sets as the problem at hand. We select the design distribution $p^*$ with support covering the range of plausible values for $\btheta_2$.  Given these values, we generate $U(\bs)$ at $\bs\in\{\bs_i,\bs_{(i)}\}$. The feature set $\bx_i$ for modeling $u_i$ at site $\bs_i$ thus contains the spatial parameters $\btheta_2$, and process values at the neighboring sites $U(\bs_{(i)})$. Since we can generate arbitrarily large datasets from the design distribution, model fit is not affected by any data scarcity of the observations. This is important since NNs often require large datasets for training. 


The NNs have their own hyperparameters which cannot be estimated directly but rather need to be tuned. These include the network architecture - the number of hidden layers ($H$), the size of each hidden layer ($L_h$), the number of basis functions ($K$), the activation function ($\psi(\cdot)$), etc. They also include NN training parameters like the learning rate, batch size, number of epochs, and early stopping criteria. We have assumed the same network architecture for all the NNs in \eqref{e:ffnn}, with the exception of differences due to a smaller Vecchia neighboring set for the first few sites. The model is fit using the \texttt{R} \citep{R} package \texttt{SPQR} \citep{SPQR_R} whose in-built cross-validation functions can be used to tune the NN hyperparameters.
Once the weights have been learned, applying the NN to the approximate likelihood is straightforward, and the Vecchia approximation ensures that the computational burden increases linearly in the number of spatial locations.
Algorithm \ref{a:local} outlines the local SPQR approximation.

\begin{algorithm}
\caption{Local SPQR approximation}\label{a:local}
\begin{algorithmic}
\Require sites $\bs_1, \ldots, \bs_n$ with sets of neighboring locations $\bs_{(1)}, \ldots, \bs_{(n)}$
\Require Design distribution $p^*$, training sample size $N$
\State $i \gets 2$
\While{$i \leq n$}
\State $j \gets 1$
\While{$j \leq N$}
    \State Draw values of $\mathbf{\btheta}_{2j} \sim p^*$
    \State Generate $U_j(\bs)$ at $\bs \in \{\bs_i, \bs_{(i)}\}$ given $\mathbf{\btheta}_{2j}$ using  \eqref{eq:copula_pareto-02}
    \State Define features $\bx_{ij} = ( \mathbf{\btheta}_{2j},u_{(i)j})$, where $u_{(i)j} = \{U_j(\bs); \bs\in\bs_{(i)}\}$
    \State $j \gets j + 1$
    \EndWhile
\State solve $\hat{{\cal W}_i} \gets \underset{{\cal W}}{\operatorname{argmax}}  \prod_{j=1}^N f_i(u_{ij}|\bx_{ij},{\cal W})$ for $f_i(u_i|\bx_i,{\cal W})$ defined in \eqref{e:1} using {\tt SPQR}
    \State $i \gets i + 1$
\EndWhile
\end{algorithmic}
\end{algorithm}

\subsection{Parameter estimation using MCMC for the NPMM}
Given the approximate model in \eqref{e:changeofvariables}--\eqref{e:vecchia} for $f_y$ with an SPQR approximation for the spatial dependence $f_u$, a Bayesian analysis using Markov Chain Monte Carlo (MCMC) methods is used for parameter estimation. We use Metropolis updates for both $\btheta_1$ and $\btheta_2$.  For an STVC model with local GEV coefficients for site $i$, we update parameters $\{\mu_t(\bs_i), \sigma(\bs_i), \xi(\bs_i)\}$ as a block sequentially by site, and exploit the Vecchia approximation to use only terms in the likelihood corresponding to sites which appear either as the response variable in the Vecchia approximation or in a Vecchia neighbor set. The coefficients $(\beta_{i0},\beta_{i1})$ are updated as a block for each $i$, and the weight parameters $\delta_{ti}, i=1,2$ are updated as a result of that. All Metropolis updates are tuned to give acceptance probabilities of 0.4, and convergence is diagnosed based on the visual inspection of the trace plots.

\section{Density Estimation for CUS Sites and Numerical Studies}\label{s:numericalstudies}
\begin{figure}
    \centering
     \includegraphics[scale=0.6]{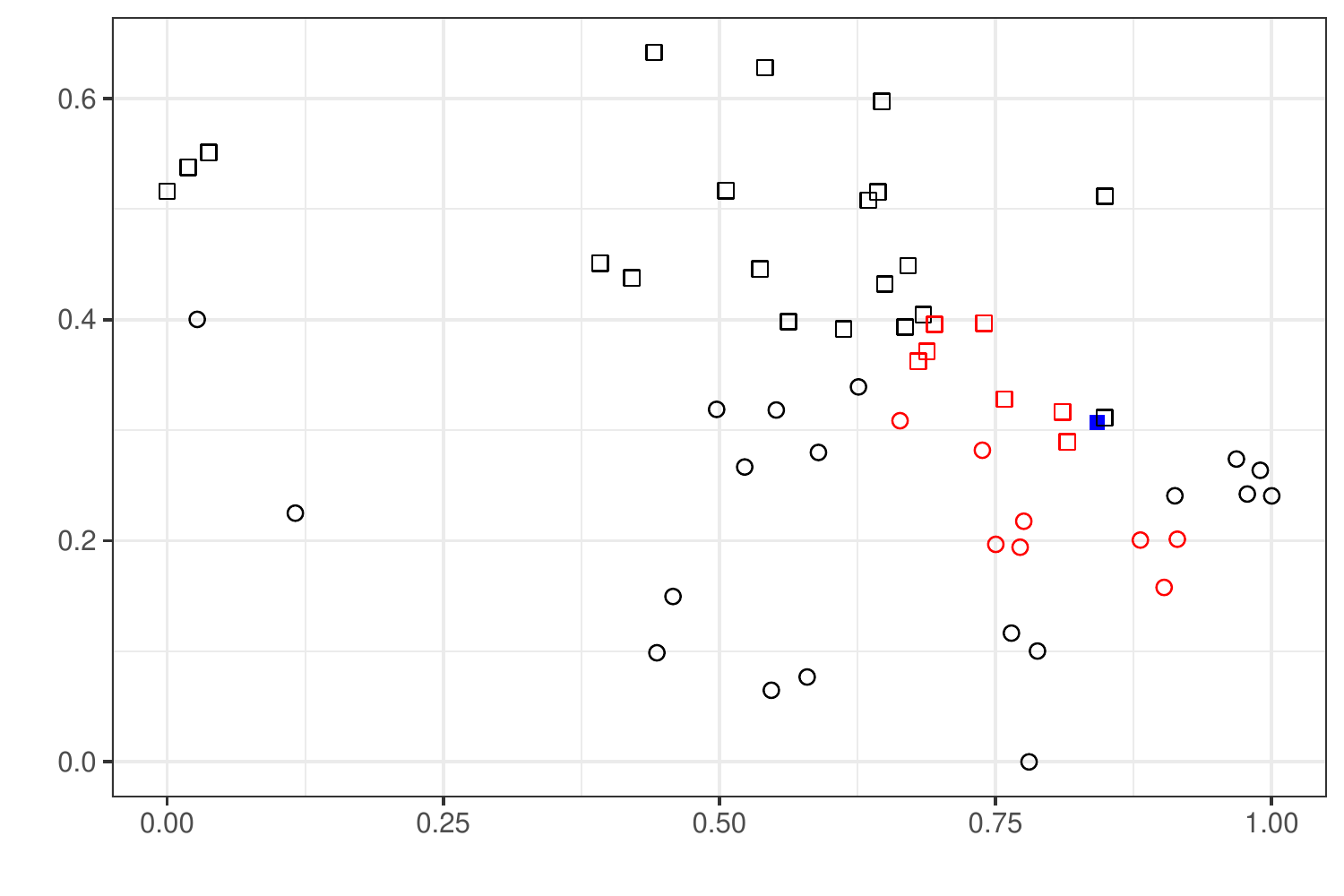}
\caption{\textbf{\small Sites used to fit SPQR models}: Distribution of 55 watershed locations scaled to the unit square. Squares and circles denote the 2 different regions. The blue square corresponds to site 45, and the red squares and circles correspond to its Vecchia neighboring set.}
    \label{fig:locations}
\end{figure}

Density estimation for the NPMM only requires knowledge of the spatial configuration of sites, and a reasonable design distribution. We consider the $n=55$ HCDN sites with the domain scaled to the unit square for convenience. Sites are assigned to the two different regions with their own weight parameters based on which HUC-02 region they belong to. Figure \ref{fig:locations} plots the distribution of the 55 sites, alongside site 45 and its Vecchia neighboring set of $m=15$ neighbors. We assume a common smoothness parameter $\alpha_R = \alpha_W = 1$ to put the 2 spatial processes on the same scale. A further assumption is made to improve model identifiability; we parameterize $\rho_W$ and $\rho_R$ to have the same effective range. We define the effective range as the distance at which the GP correlation reaches $0.05$ and the extremal coefficient $\chi$ for the MSP reaches $0.05$. In \citep{MajumderReichShaby2022}, this was achieved by setting $\rho = \rho_W$ and $\rho_R = 0.19\rho$. 
\begin{figure}
    \centering
    \begin{subfigure}{0.49\linewidth}
    \includegraphics[scale=0.55]{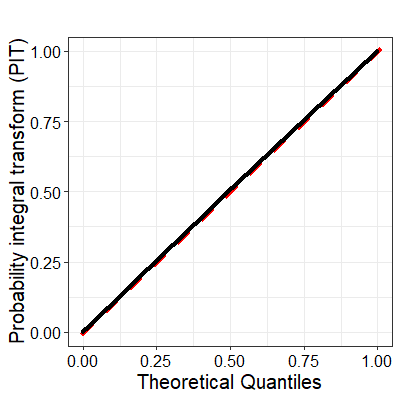}
\caption{\small Q-Q plots for goodness of fit.}
    \label{fig:qqplot_sim}
    \end{subfigure}
    \hfill
    \begin{subfigure}{0.49\linewidth}
    \includegraphics[scale=0.55]{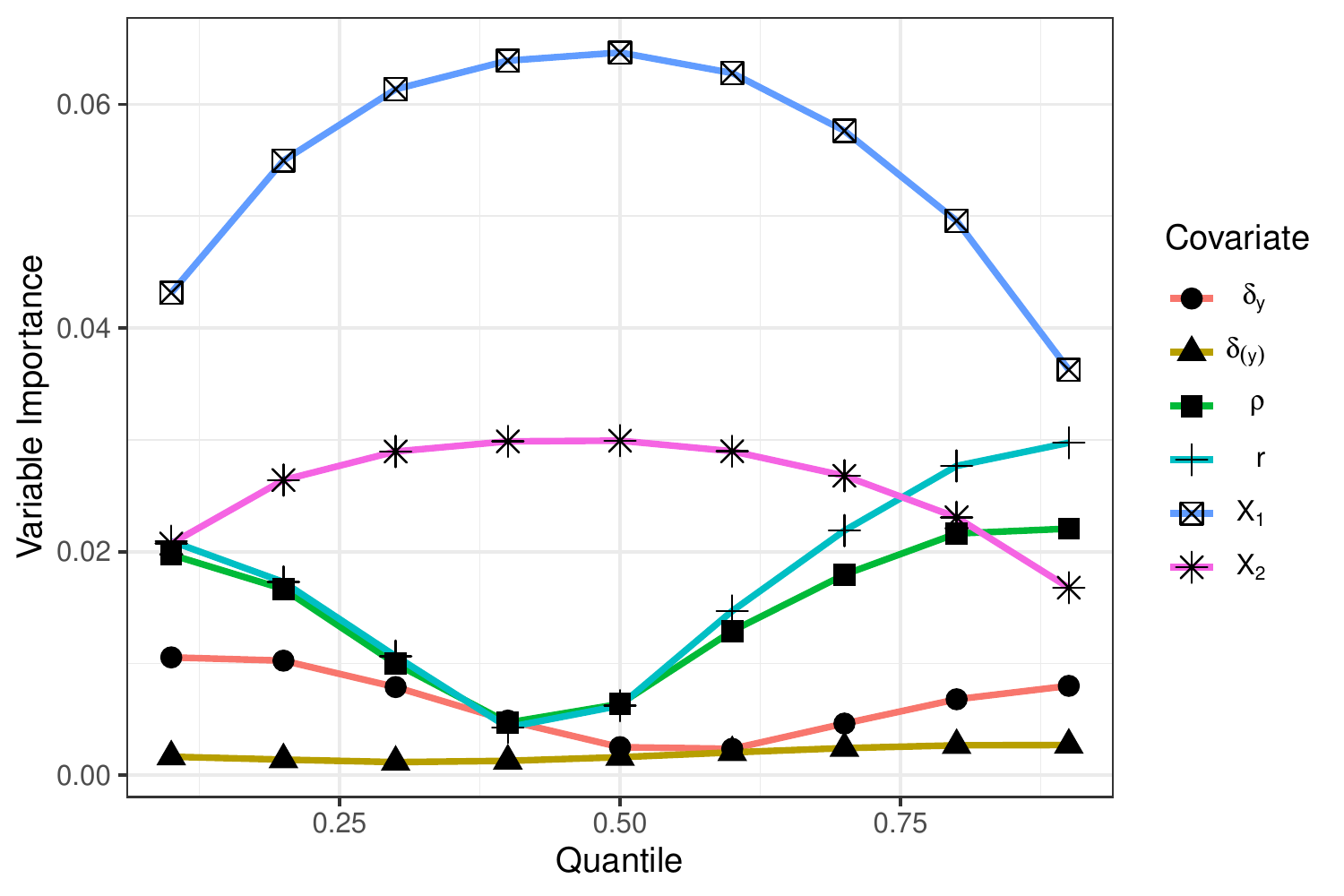}
\caption{\small Variable importances.}
    \label{fig:VI}
    \end{subfigure}
    \caption{\small {\bf Model diagnostics for local SPQR fit at site 45}: Q-Q plot (left) for goodness of fit and variable importance plot (right) for the local SPQR model. $\delta_{12}$ in the variable importance plots is defined as $\log \delta_1 - \log \delta_2$.}
    \label{fig:diagnostics}
\end{figure}

\paragraph{Local SPQR model architecture:} 
For density estimation, we fit local SPQR models for each site $\bs_i, i=2:55$. The local SPQR models have identical architectures for each site with 2 hidden layers with 30 and 20 neurons respectively, 15 output nodes, a learning rate of 0.01, and 100 epochs with a batch size of 1000. The model architecture was chosen by comparing the log-likelihood of fitted models with different architectures, and are very similar to those used in \citep{MajumderReichShaby2022}. The number of output nodes in this case correspond to the number of basis functions used to approximate the true conditional density. While the analytical form of the conditional densities are not available for the NPMM, \cite{MajumderReichShaby2022} was able to study this for a GP, which is equivalent to setting $\delta_{1t} = \delta_{2t} = 0$. The conditional densities are univariate Gaussian and analytically available in this case; 10--15 output nodes were found to be sufficient in modeling the conditional density, with higher values leading to random fluctuations in the estimated approximated conditional density. We train the SPQR models with the design distribution $p^*$, generating $2\times 10^6$ samples uniformly from $\rho,\delta_{1t},\delta_{2t},r \in (0,1)$ with all parameters independent of each other. Choosing $p^*\sim U(0,1)$ for each of the parameters allows us to explore the parameter space uniformly within its support. The response $u_i$ is a function of exactly one of $\delta_{1t}$ or $\delta_{2t}$ depending on which region $\bs_i$ belongs to. The other weight parameter is relevant for density estimation only if one of the neighbors is in the other region. Thus, some sites require exactly one of $\delta_{1t}$ or $\delta_{2t}$, while other sites require both. To ensure consistent dimensions of the feature vector across locations as well as identifiability of the weight parameters, we define $\delta_y$ and $\delta_{y'}$ to be the weight parameters corresponding to the response and the neighbors respectively. If all neighbors belong to the same region as the response, $\delta_{y'} = \delta_y$. Finally, we define $\delta_{(y)} = \log \delta_y - \log \delta_{y'}$, which is non-zero only if some of the neighbors belong to a different region from the response. Instead of using $\delta_{1t}$ and $\delta_{2t}$, we use $\delta_y$ and $\delta_{(y)}$ as covariates for density estimation.  Algorithm \ref{a:local} is then used to fit the local SPQR models.

Figure \ref{fig:qqplot_sim} plots the probability integral scores for the local SPQR model at site 45; the scores falling along the $Y=X$ line (partially visible, in red) suggests a good model fit. Figure \ref{fig:VI} plots the variable importance scores for the two nearest neighbors (denoted as $X_1$ and $X_2$) as well as the spatial parameters of the process. The neighbors have the highest importance across the quantiles, and the spatial parameters are important covariates for at least one of the extremal quantiles. The remaining neighbors have significantly lower importances compared to the first few and have been omitted from the plot for clarity; their exact magnitude often depends on the spatial configuration of the locations. Variable importance plots for additional locations are provided in \ref{s:VI-plots}.

\begin{table}
\centering
\caption{\small True parameter values for the 3 simulation study scenarios.}
\label{t:sim}
\begin{tabular}{cccccccccc}
\toprule
Scenario & $\mu_0$ & $\mu_1$ & $\sigma$ & $\xi$ & $\rho$ & $\beta_{10}$ & $\beta_{11}$ & $\beta_{20}$ & $\beta_{21}$ \\\midrule
1 & 12 & 3 & 2 & 0.2 & 0.4 & -1 & 1.8 & 0.2 & 2 \\
2 & 13 & 5 & 2 & 0.1 & 0.1 & 1 & -1.2 & -1 & 0.8\\
3 & 12 & 3 & 3 & -0.1 & 0.2 & -1.5 & 2 & -1.5 & 0.8 \\\bottomrule
\end{tabular}
\end{table}

\paragraph{Numerical study for parameter estimation:} Before using the density estimates on the observed annual streamflow maxima data, we consider 3 scenarios with different spatial and marginal GEV parameters in order to ascertain how the density-estimation errors propagate to parameter-estimation errors. We assume $\delta_{1t}$ and $\delta_{2t}$ are independent of each other and depend on time by means of a probit link function, i.e.,
\begin{align}\label{e:delta}
    \Phi^{-1}(\delta_{it}) = \beta_{i0} + \beta_{i1} Z_{it}, i = 1,2.
\end{align}
As covariates, we use $Z_{1t} = (t - \Bar{t})/10$ and $Z_{2t} = Z_{1t} - 0.05$, where $t = 1972 + t - 1$ and $\Bar{t}$ is the mean of $t$. For all cases, the location parameters of the GEV are assumed to depend on a covariate as in \eqref{e:marginal}, and we use $X_t(\bs) = Z_{1t}$ for all sites. Within a scenario, each site is assumed to have the same marginal GEV parameters. Table \ref{t:sim} lists the true parameter values for the 3 scenarios.

We generated 60 datasets for each scenario. Each dataset contains 50 independent realizations of the NPMM at the 55 sites shown in Figure \ref{fig:locations}. For priors, we select $\mu_0, \mu_1, \log(\sigma)\sim \mbox{Normal}(0,10^2)$, $\xi \sim \mbox{Normal}(0,0.25^2)$, $\beta_{10},\beta_{11},\beta_{20},\beta_{21}\sim \mbox{Normal}(0,1)$, and $\rho, r \sim \mbox{Uniform}(0,1)$. We approximate the posterior using MCMC with 11,000 iterations and Metropolis candidate distributions tuned to have an acceptance probability of around 0.4. After discarding the first 1,000 iterations as burn-in, we compute posterior means and 95\% confidence intervals for each parameter based on the remaining samples. The posterior distributions of $\beta_{10},\beta_{11},\beta_{20},\mbox{ and }\beta_{21}$ are used to evaluate the posterior distributions of the mean of $\delta_{1t}$ and $\delta_{2t}$.
\begin{figure}
    \centering
     \includegraphics[width=\linewidth]{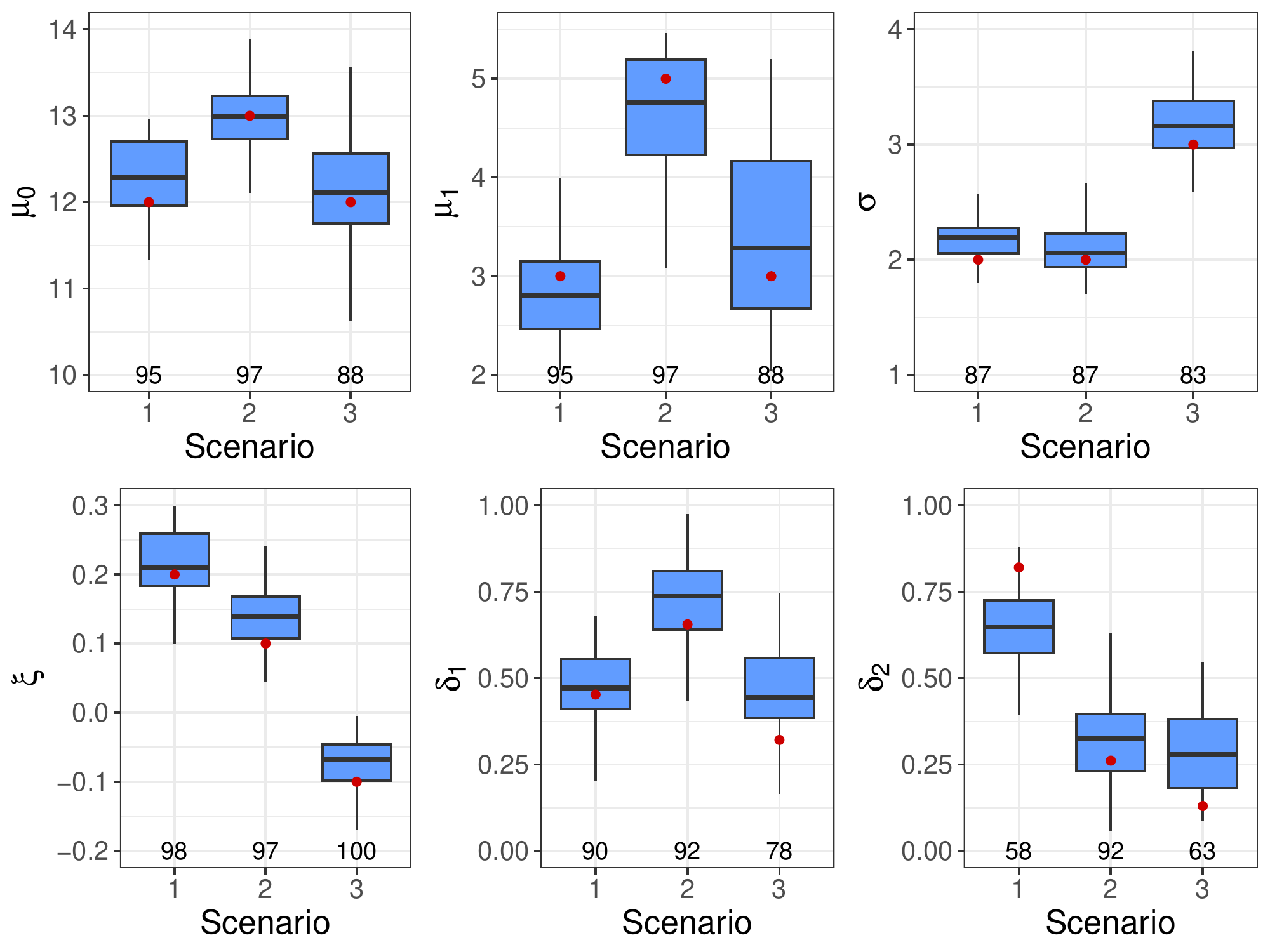}
        \caption{\small \textbf{Marginal and spatial parameter estimates:} Sampling distribution of the posterior mean for GEV and spatial parameters for the three simulation scenarios. The red dots are the true values, and empirical coverage of the 95\% intervals are provided at the bottom of each plot.}
    \label{fig:sim_boxplot}
\end{figure}

Figure \ref{fig:sim_boxplot} plots the sampling distribution of the posterior mean estimator of model parameters of interest and provides the empirical coverage of $95\%$ posterior intervals at the bottom of each panel. Posterior estimators of the GEV parameters have relatively little bias and nominal coverage. To evaluate the posterior of $\delta_{1t}$ and $\delta_{2t}$, we plot $\delta_i = \frac{1}{50}\sum_{t=1}^{50}\delta_{it}$, for $i = 1,2$. Estimation of $\delta_i$ proves more challenging, likely due to the spatial configuration of the locations, and the relatively low importance of $\delta_{y}$ and $\delta_{(y)}$ in the SPQR model. While bias and variability are higher for the spatial parameters, but our methods can still distinguish between the asymptotic regimes of $\delta_{1}$ and $\delta_{2}$.

\section{Analysis of Extremal Streamflow in Central US}
\subsection{Model description}\label{s:model_description}
We assign an STVC model to each of the marginal GEV parameters. The responses are modeled as 
  \begin{equation}
      Y_t(\bs) \sim \mbox{GEV}\left\{\mu_0(\bs) + \sum_{j=1}^5 \mu_j(\bs)X_{jt}(\bs),\sigma(\bs), \xi(\bs) \right\}.
\end{equation}
The intercept process $\mu_0(\bs)$ is assigned a GP prior with nugget effects to allow local heterogeneity:
\begin{align*}
      \mu_0(\bs) &= \tilde{\mu}_0(\bs) + e_0(\bs) \\
      e_0(\bs)&\iid \mbox{Normal}(0,v_{\mu_0})\\
      \tilde{\mu_0}(\bs)&\sim \mbox{GP}(\beta_{\mu_0},\tau^2_{\mu_0}K(\bs,\bs')), \mbox{ where } K(\bs,\bs') = \exp\{-||\bs-\bs'||/\rho_{\mu_0}\}\\
      \beta_{\mu_0} &\sim \mbox{Normal}(0,10^2), \tau^2_{\mu_0}, v^2_{\mu_0}\iid \mbox{IG}(0.1,0.1), \log \rho_{\mu_0}\sim \mbox{Normal}(-2,1),
  \end{align*}
where IG$(\cdot,\cdot)$ is the inverse-Gamma distribution. The slopes $\mu_j(\bs)$, $j=1:5$, the log-scale $\log \sigma(\bs)$, and the shape $\xi(\bs)$ are modeled similarly using GPs. The STVC parameters are denoted as $\btheta_3 = \{\beta_{\mu_0},\tau^2_{\mu_0},\rho_{\mu_0},\ldots,\beta_{\xi},\tau^2_{\xi},\rho_{\xi}\}$.

\begin{figure}
    \centering
    \begin{subfigure}[b]{0.48\linewidth}
     \includegraphics[width=\linewidth]{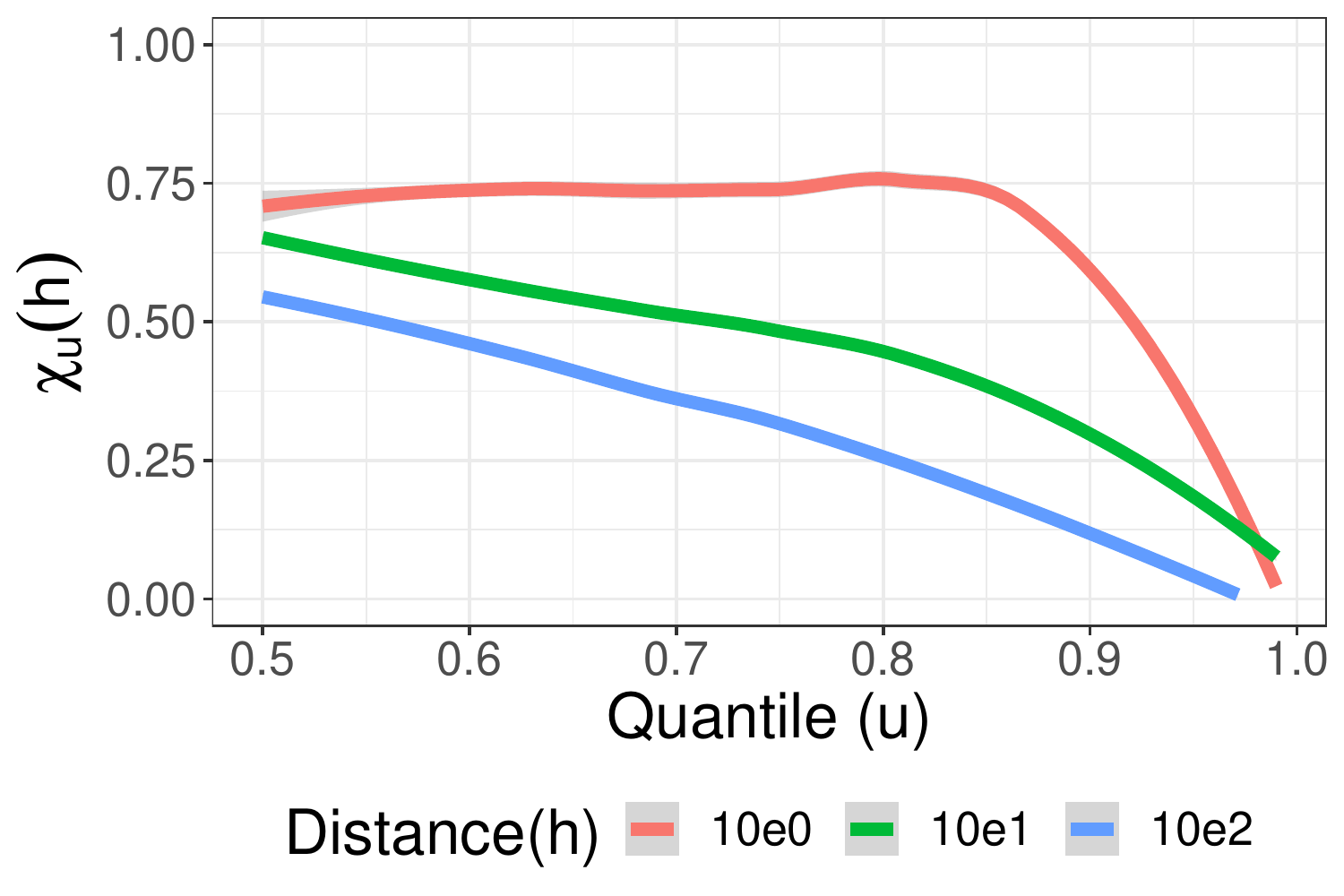}
\caption{\small Conditional exceedance $\chi_u(h)$ for annual maximum streamflow computed for different distances.}
    \label{fig:chi_h}
    \end{subfigure}
    \hfill
    \begin{subfigure}[b]{0.48\linewidth}
        \ \includegraphics[width=\linewidth]{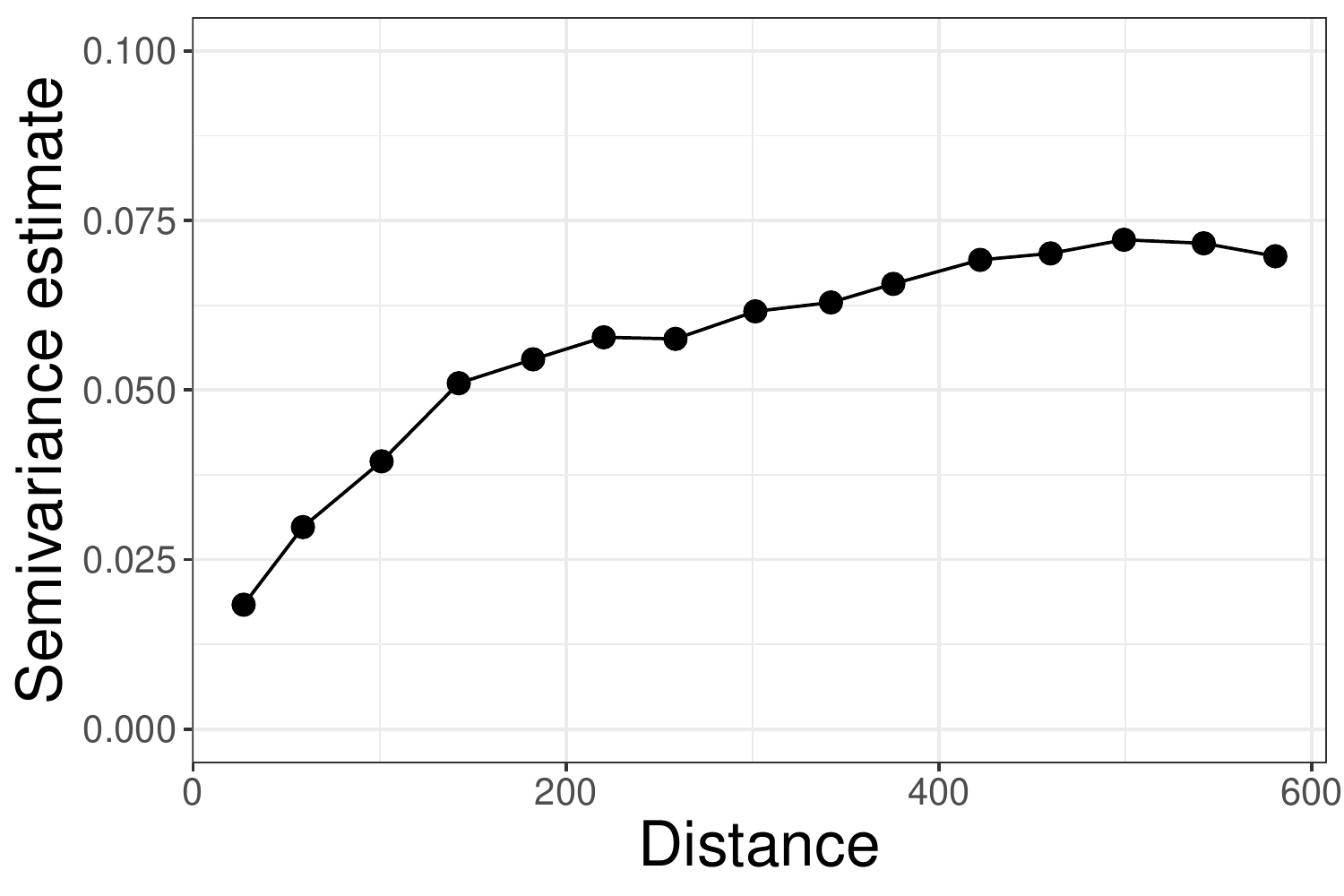}
\caption{\small Sample variogram for annual maximum streamflow, averaged over 50 years of data.}
    \label{fig:avg_variogram}
    \end{subfigure}
        \caption{\small Spatial behavior of annual maximum streamflow in terms of the conditional exceedance and the variogram.}
    \label{fig:spatial_HUC02}
\end{figure}

For the residual model, we use the process mixture model in Section \ref{s:model} for spatial dependence and assume independence across years. The simplifying assumptions that we make for the MSP $R_t(\bs)$ and the GP $W_t(\bs)$ in Section \ref{s:numericalstudies} are maintained here. The model for the weight parameters $\delta_{1t}$ and $\delta_{2t}$ along with the priors for all parameters in $\btheta_2$ are written as:
  \begin{align*}
  \Phi^{-1}(\delta_{it}) &= \beta_{i0} + \beta_{i1} Z_{it}, i = 1,2\\
   \beta_{10},\beta_{11},\beta_{20},\beta_{21} &\sim \mbox{Normal}(0,1)\\
   \rho,r &\sim \mbox{Uniform}(0,1).
  \end{align*}
  Note that the priors on the spatial ranges are for the scaled domain. In addition, both the streamflow and precipitation data have been rescaled to [0,1] to ensure stable estimates. Figure \ref{fig:chi_h} plots $\chi_u(h)$ for rank-standardized streamflow data as a function of $u$ for different values of $h$. The rank standardization ensures a Uniform$(0,1)$ marginal distribution at each location. The plot suggests an asymptotically independent process.
Figure \ref{fig:avg_variogram} plots the mean of the annual variograms of the streamflow data. It shows a range of over 500 km, as well as the presence of a nugget effect. 
\subsection{Extremal streamflow patterns within the CUS}

The local SPQR models from Section \ref{s:numericalstudies} are used to compute the density estimates. For parameter estimation, we ran 2 independent MCMC chains for 15,000 iterations each, discarding the first 5,000 of each chain as burn-in. Table \ref{t:posterior} lists the posterior means and standard deviations of the spatial parameters based on the 20,000 post-burn-in posterior samples.
\begin{table}
\centering
\caption{\small Posterior means and standard deviations (SD) of spatial parameters of the NPMM based on MCMC.}
\label{t:posterior}
\begin{tabular}{ccc|ccc}
\toprule
\textbf{Parameter} & \textbf{Mean} & \textbf{SD} & \textbf{Parameter} & \textbf{Mean} & \textbf{SD} \\\midrule
$\beta_{10}$ & -0.15 & 0.33 & $\rho$ & 0.25 & 0.49 \\
$\beta_{11}$ & 0.92 & 0.86 & $r$ & 0.88 & 0.03 \\
$\beta_{20}$ & 0.47 & 0.41 & $\delta_{1}$ & 0.53 & 0.10 \\
$\beta_{21}$ & 0.65 & 0.94 & $\delta_2$ & 0.71 & 0.12 \\\bottomrule
\end{tabular}
\end{table}
The posterior mean of $r$ suggests the presence of a nugget effect. For the posterior distribution of $\delta_i, i=1,2$, we evaluate $\frac{1}{50}\sum_{t=1}^{50} \delta_{it}$ for each posterior MCMC sample of $(\beta_{i0},\beta_{i1})$ and interpret it as the average value of the weight parameter conditioned on precipitation. The empirical $95\%$ confidence intervals for the slope parameters $\beta_{i1}$ are $\beta_{11} \in (-0.76,2.56)$, and $\beta_{21} \in (-1.18,2.47)$; both intervals include zero, suggesting that the weight parameters for the two regions which ascribe  the asymptotic regime of extremal streamflow are not associated with the annual regional precipitation. 
\begin{figure}
    \centering
    \includegraphics[scale=0.8]{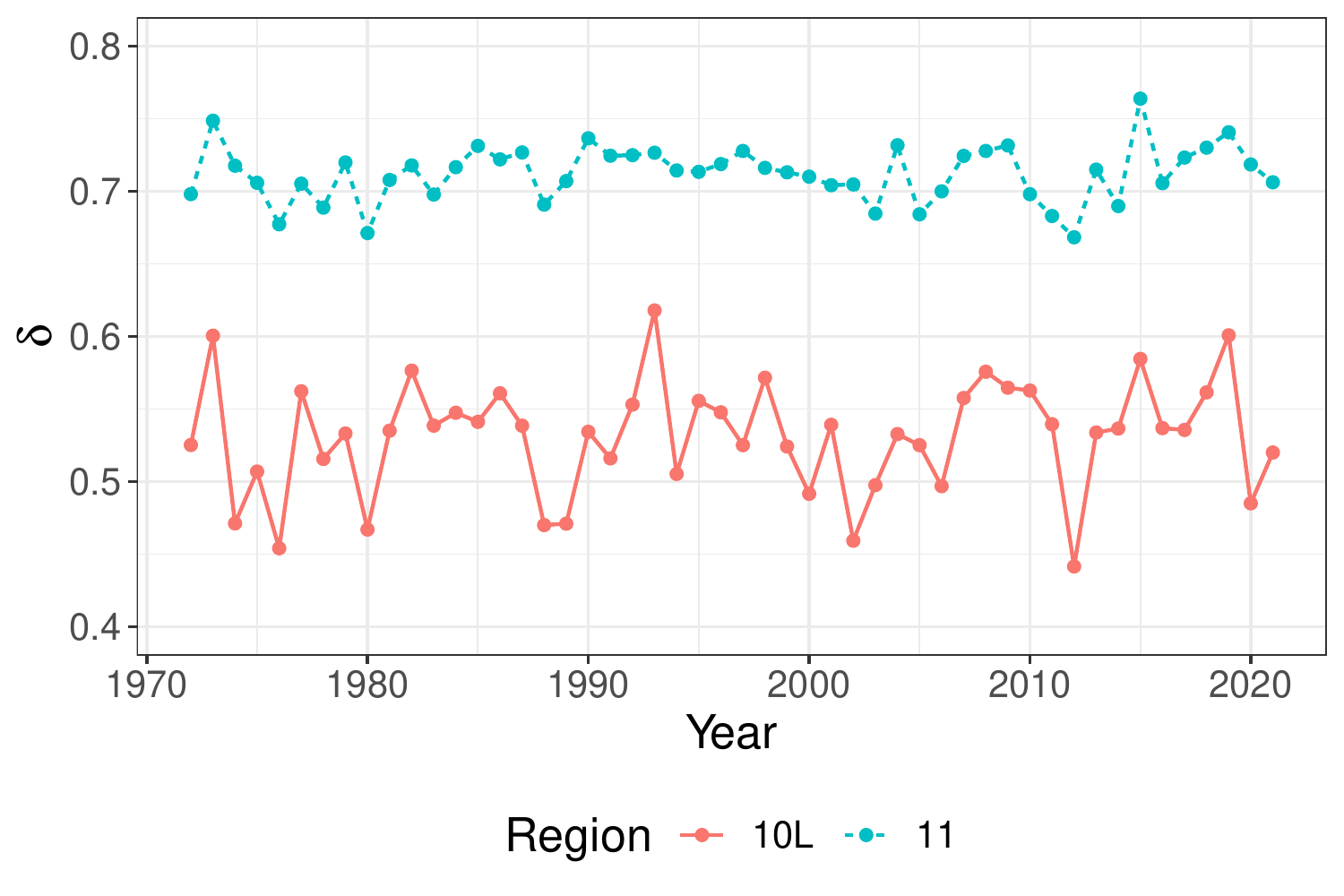}
\caption{\small Posterior means of $\delta_1$ corresponding to region 10L and $\delta_2$ corresponding to region 11, computed annually for 1972--2021.}
    \label{fig:delta_by_year}
\end{figure}

To understand changes in $\delta_{it}$ as a function of annual precipitation, we evaluate it for 1972--2021 based on the posterior means of $(\beta_{i0},\beta_{i1})$.  Figure \ref{fig:delta_by_year} plots the value of the weight parameter for the 2 HUC-02 regions from 1972--2021. Region 11 which corresponds to the lower half of the CUS, has a higher estimate of the weight parameter than region 10L. The sites in region 11 tend to show asymptotic dependence, while the sites in region 10L vary between asymptotic dependence and asymptotic dependence in different years. The estimates are quite different for the 2 regions and vary quite a lot from year to year for region 10L, indicating the appropriateness of the non-stationarity assumption of the spatial process.

\begin{figure}
    \centering
    \begin{subfigure}{0.49\linewidth}
    \includegraphics[scale=0.55]{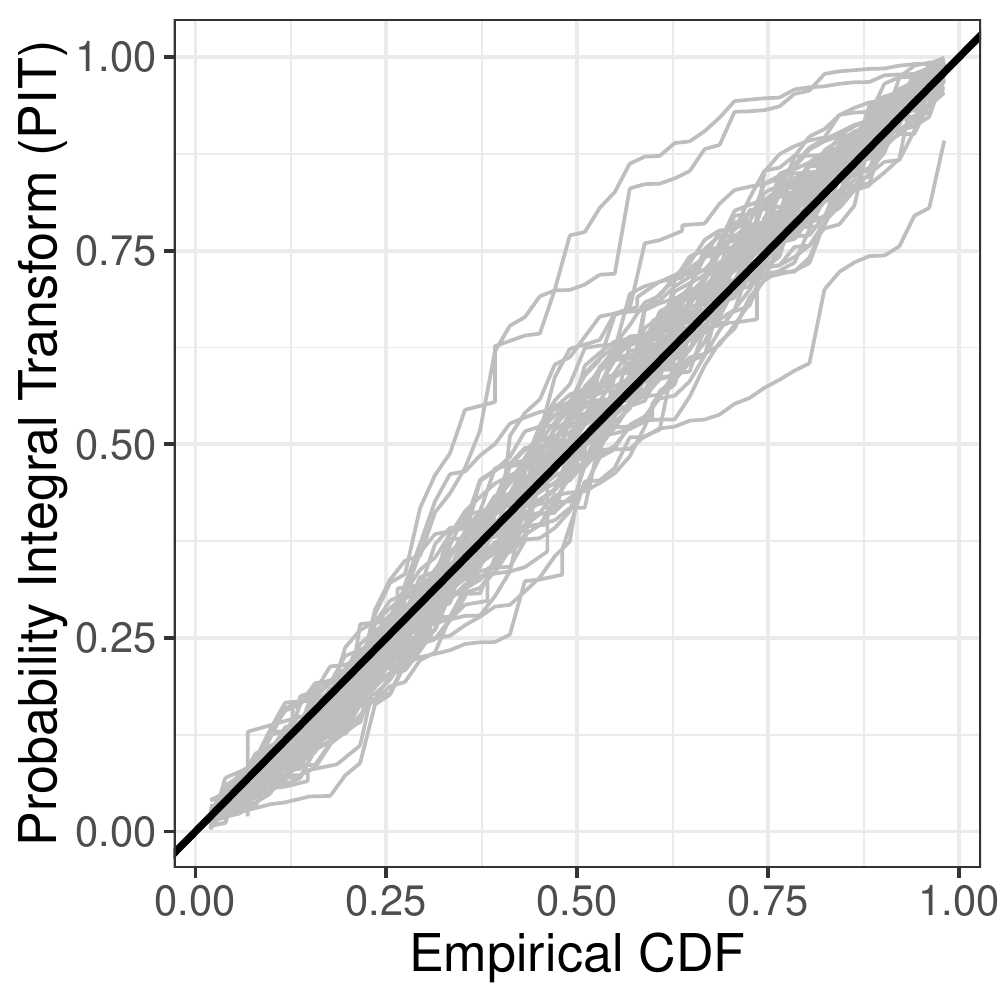}
\caption{\small Q-Q plots based on independent MLE estimates.}
    \label{fig:fit_MLE}
    \end{subfigure}
    \hfill
    \begin{subfigure}{0.49\linewidth}
    \includegraphics[scale=0.55]{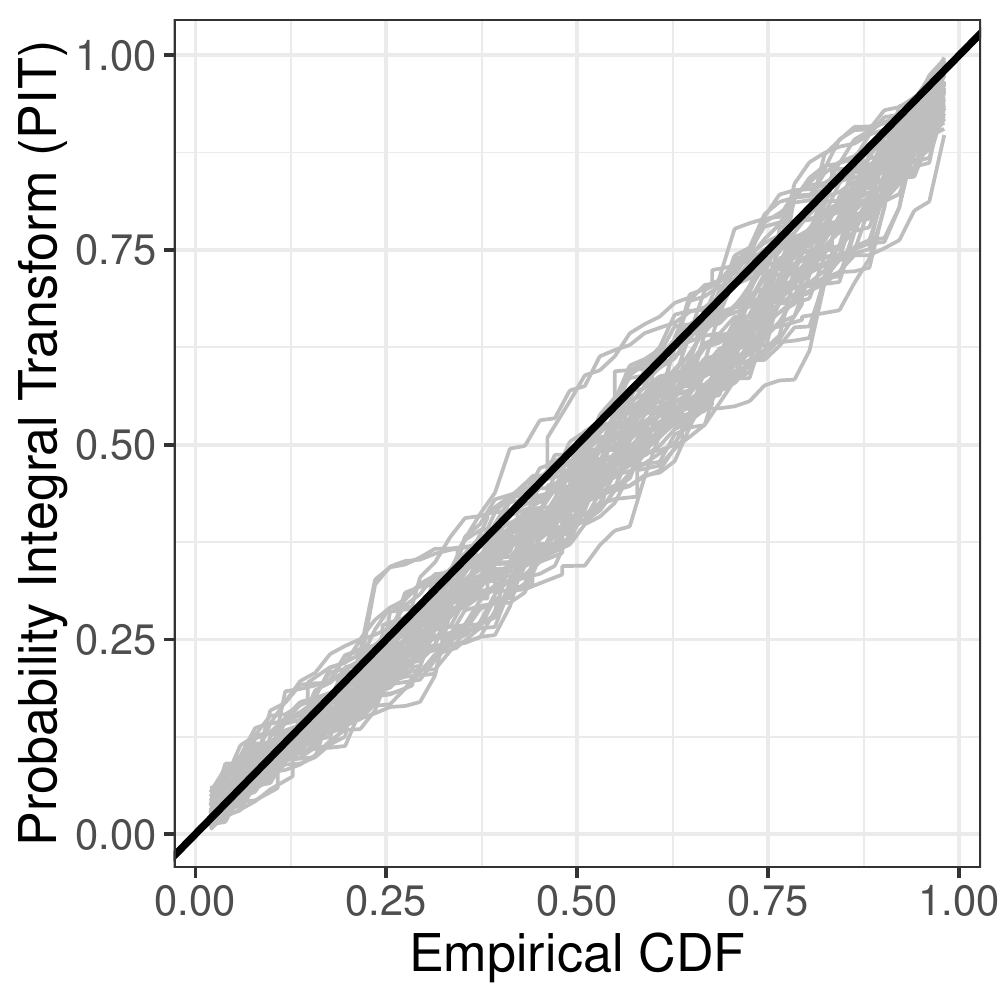}
\caption{\small Q-Q plots based on NPMM posterior means.}
    \label{fig:fit_NPMM}
    \end{subfigure}
    \caption{\small {\bf Goodness of fit for the marginal distributions of annual streamflow maxima}: Q-Q plots for MLE computed independently at all sites (left), and based on posterior means from the NPMM (right).}
    \label{fig:qqplot}
\end{figure}

Figure \ref{fig:qqplot} shows the goodness of fit of the marginal GEV models, based on maximum likelihood estimates (MLE) computed individually at each site in \ref{fig:fit_MLE}, and estimates derived using the posterior means of the NPMM in \ref{fig:fit_NPMM}. Visual inspection suggests that the NPMM provides overall better fits compared to independent MLE despite having more bias. We compared the standard errors of the GEV parameters based on the MLE with the posterior standard deviation of the GEV parameters based on the NPMM, and found that the latter was always lower; see Table \ref{t:GEV_fits} in \ref{s:marg-estimates} for more details. Since extremes data is often scarce by definition, pooling in spatial information across sites is crucial for improving model fits and in turn getting valid inference. The posterior means and standard deviations for the components of $\btheta_3$ are also provided in Table \ref{t:SVC_params}.

\begin{figure}
    \centering
    \begin{subfigure}{0.49\linewidth}
    \includegraphics[scale=0.45]{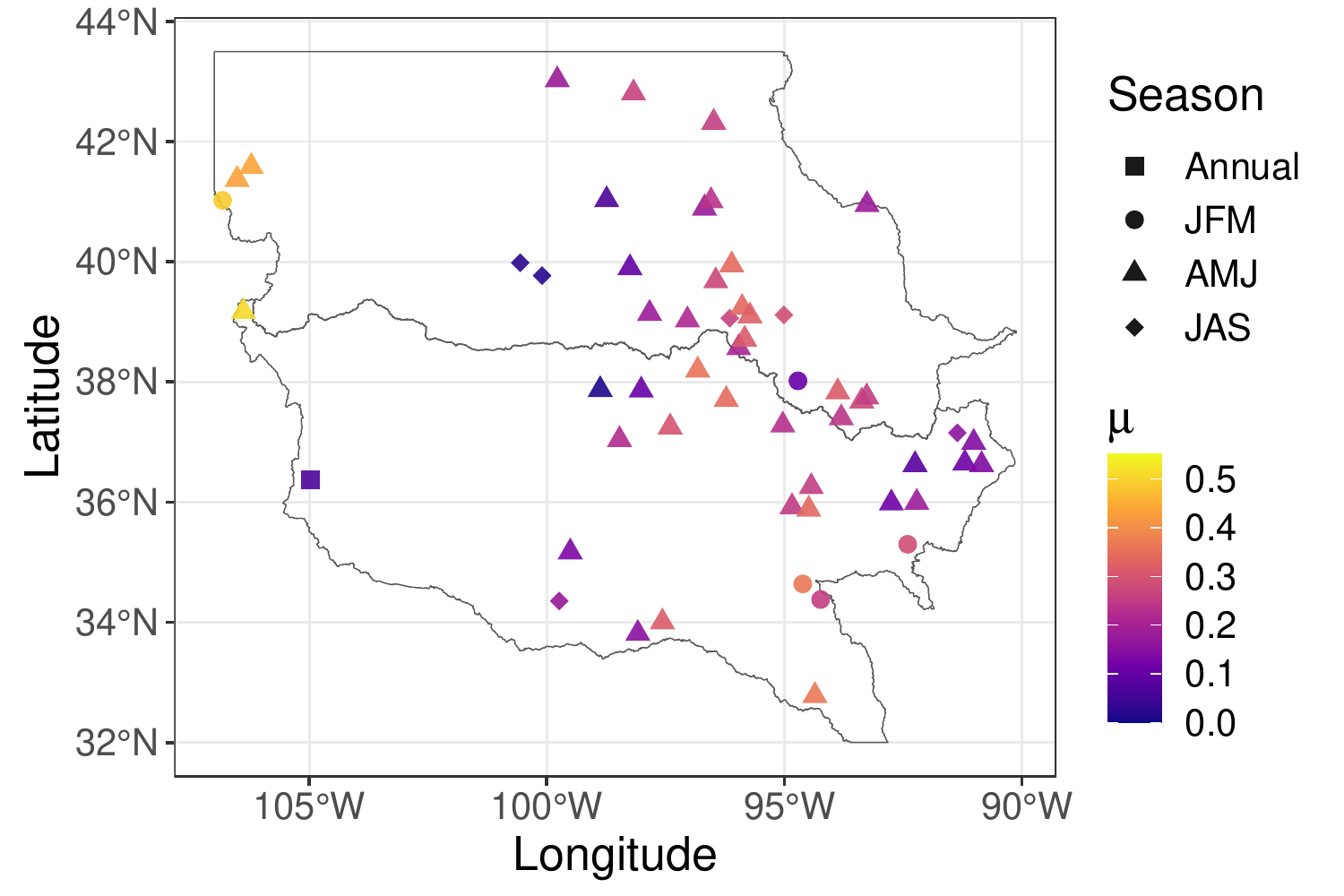}
\caption{\small Estimates of largest slope parameter $\mu(\bs)$ at each site.}
    \label{fig:fitted_location}
    \end{subfigure}
    \hfill
    \begin{subfigure}{0.49\linewidth}
    \includegraphics[scale=0.45]{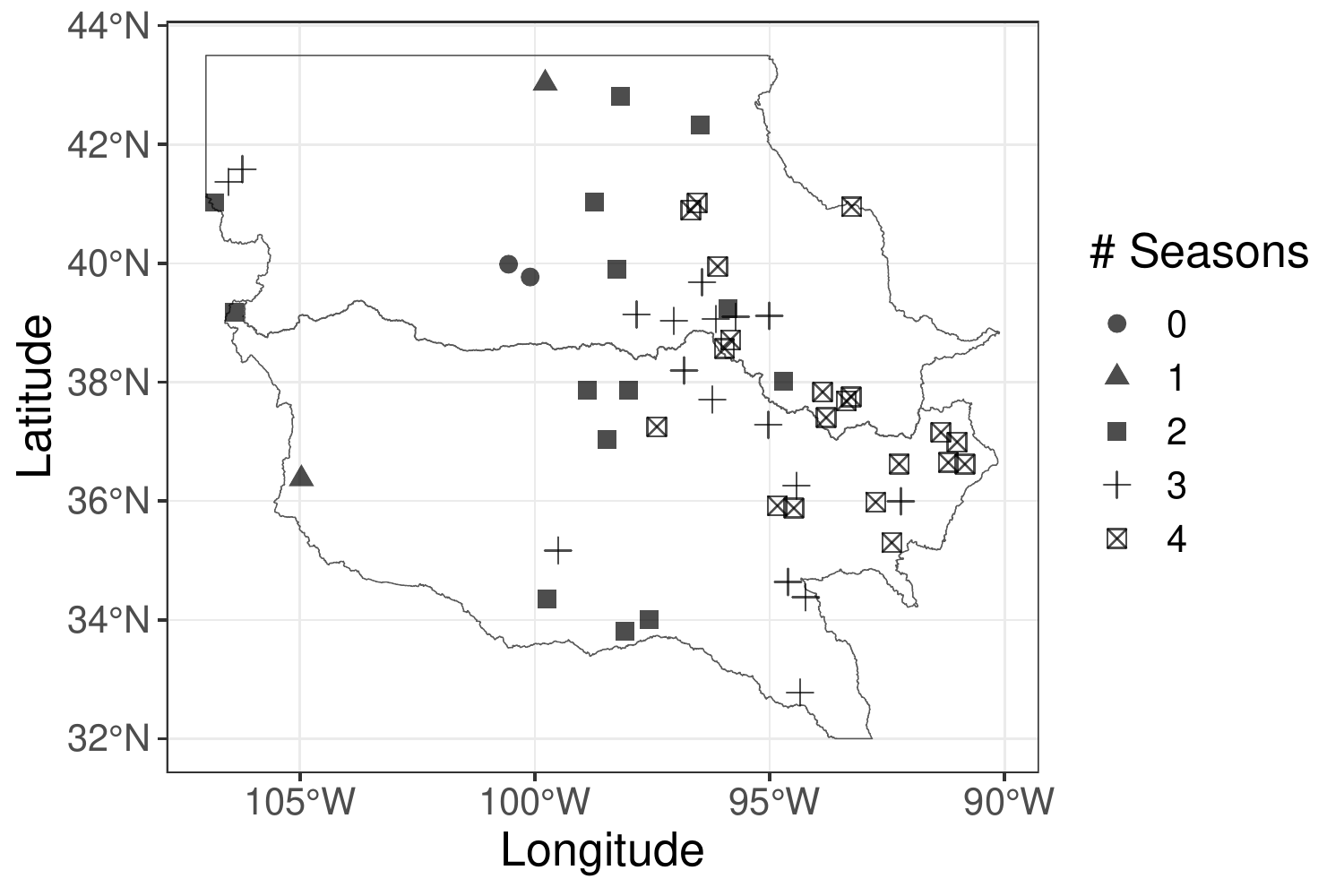}
\caption{\small Number of seasons with for which $\mu(\bs)>0$ with high probability.}
    \label{fig:fitted_seasons}
    \end{subfigure}
    \caption{\small {\bf Posterior means of slope parameters for annual streamflow maxima}: Estimates of $\mu(\bs) = \max(\mu_j(\bs))$ for $i=j(1)4$ corresponding to the 4 seasons with shapes denoting the season with the highest slope value (left), and number of seasons (excluding annual) where $\mathbb{P}[\mu(\bs)>0]>0.90$ (right).}
    \label{fig:estimates1}
\end{figure}

Figure \ref{fig:estimates1} shows the posterior means of the slope parameters for each HCDN site. Since each site has 5 slopes corresponding to the annual precipitation as well as 4 seasonal precipitations, we focus on the largest slope parameters for each site, corresponding to the season where precipitation has the most significant effect on streamflow. Figure \ref{fig:fitted_location} plots the slope parameter for the most significant season at each site; the colors denote the magnitudes of the slope parameter for the most significant season and the shapes denote the season it corresponds to. We see that most of the points are for spring (AMJ), and exactly one location (in region 11) is affected more by annual precipitation than by seasonal precipitation. To assess the strength of the significance for all seasons, we computed the posterior probability of each slope parameter being greater than 0, i.e., $\mathbb{P}[\mu_j(\bs)>0]$ for $j = 2:5$. The slope corresponding to the annual precipitation is not considered in this case, and all 55 sites had at least one seasonal slope with a non-zero probability. we count the number of seasons where $\mathbb{P}[\mu_j(\bs)>0] > 0.90$ for each site; the resulting plot is presented in Figure \ref{fig:fitted_seasons}. The lower values in the plot indicate that precipitation has a large effect on streamflow only in specific seasons, whereas the higher values signify that maximum streamflow is a function of seasonal precipitation from different seasons for different years. We refer the reader to \citep{Awasthietal2022} for further discussion on the seasonal/annual effect of precipitation on streamflow for different regions. Considering that most of these sites have 3--4 significant seasons as shown in Figure \ref{fig:fitted_seasons}, it is reasonable to conclude that maximum streamflow is affected by the convective storms that occur in the CUS and the associated precipitation.

\begin{figure}
    \centering
    \begin{subfigure}{0.49\linewidth}
    \includegraphics[scale=0.45]{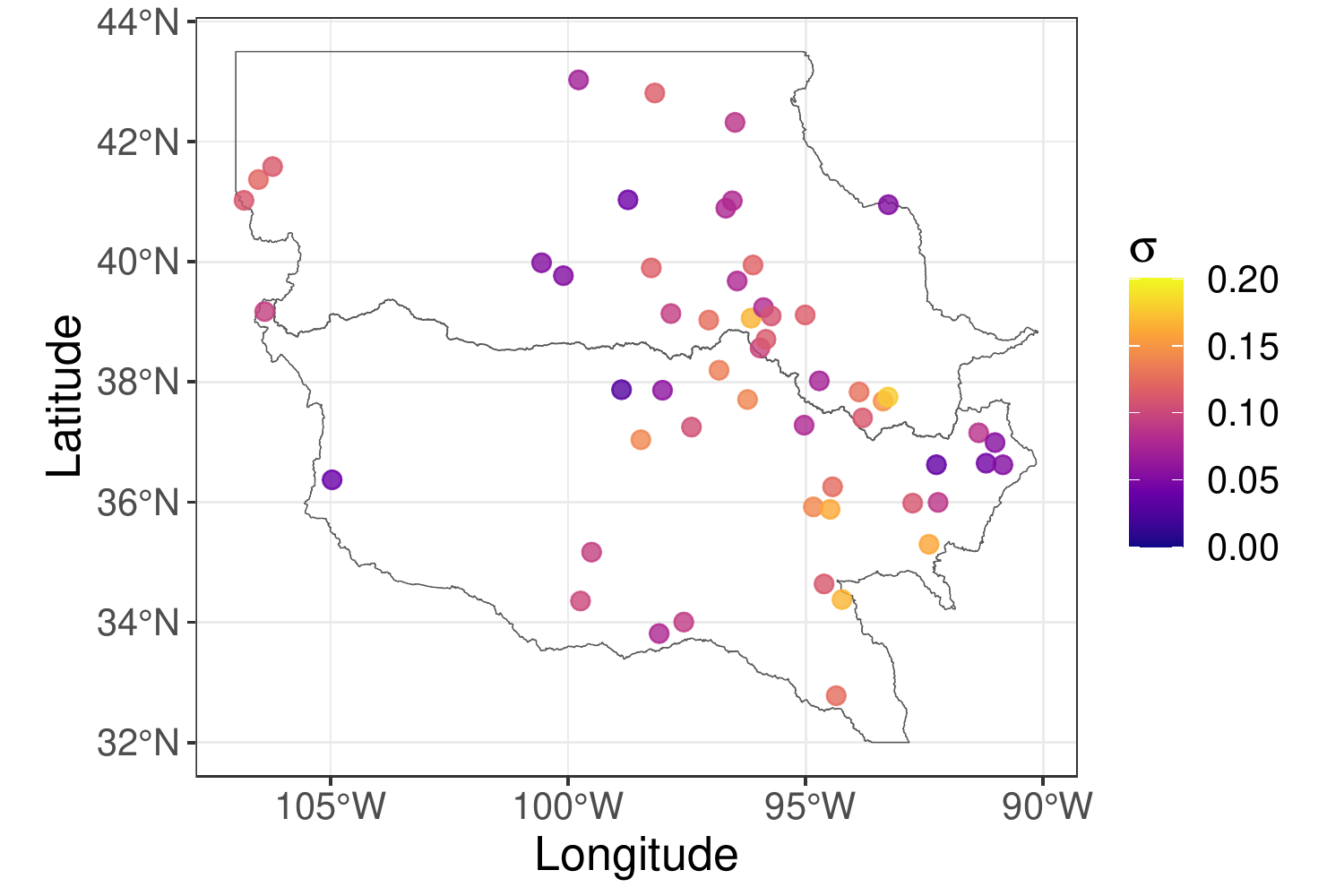}
\caption{\small Estimates of scale parameter $\sigma(\bs)$.}
    \label{fig:fitted_scale}
    \end{subfigure}
    \hfill
    \begin{subfigure}{0.49\linewidth}
    \includegraphics[scale=0.45]{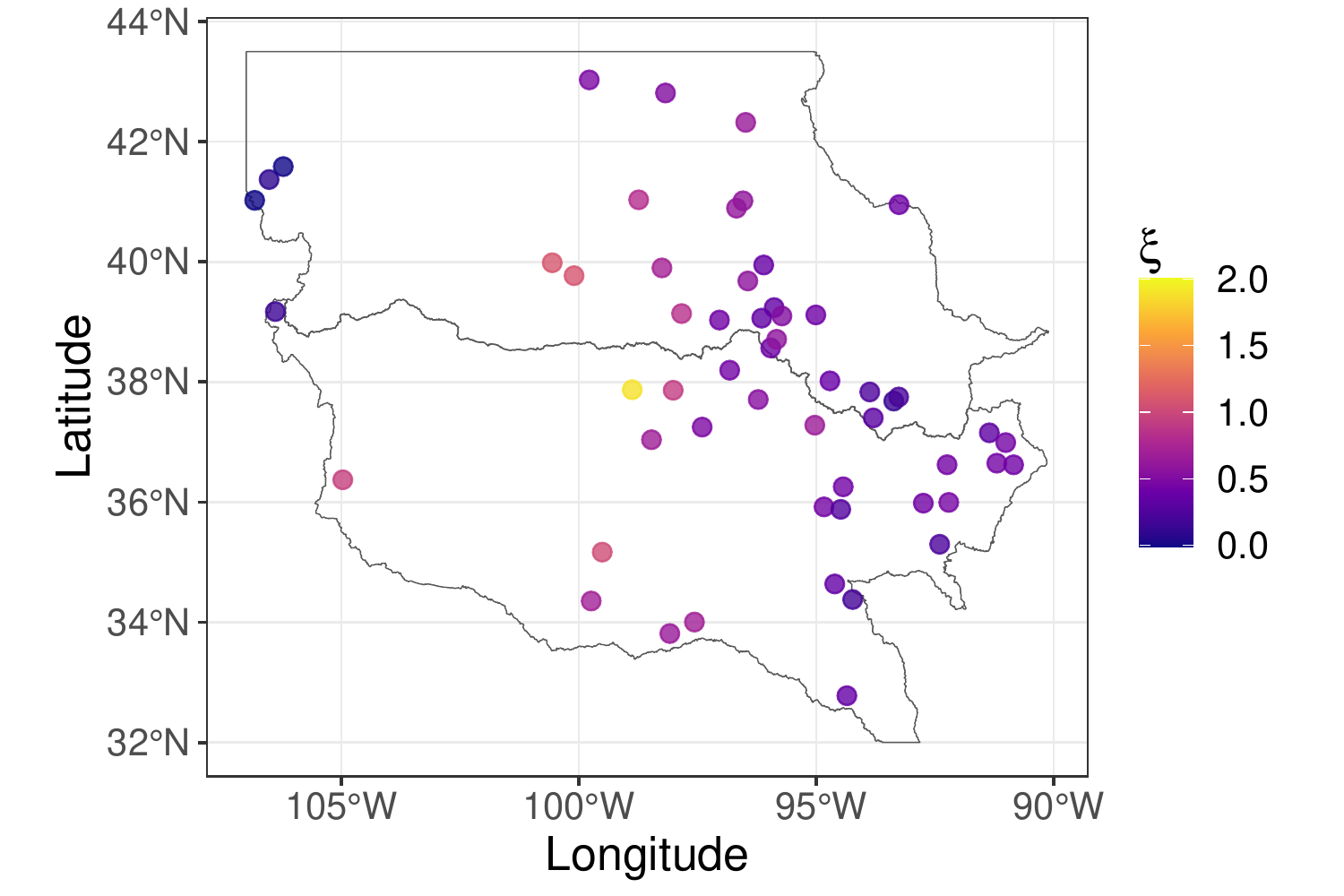}
\caption{\small Estimates of shape parameter $\xi(\bs)$.}
    \label{fig:fitted_xi}
    \end{subfigure}
    \caption{\small Posterior means of scale and shape parameters of annual streamflow maxima.}
    \label{fig:estimates2}
\end{figure}

Finally, Figure \ref{fig:estimates2} contains posterior means of the scale and shape parameters of all the watersheds. Both parameters are spatially dependent over the CUS region. We also note that the posterior means of the shape parameter are positive for 54 of the 55 sites.
\subsection{Annual streamflow maxima projections under RCP 4.5 and RCP 8.5}
\begin{figure}
    \centering
    \begin{subfigure}{\linewidth}
    \includegraphics[scale=0.8]{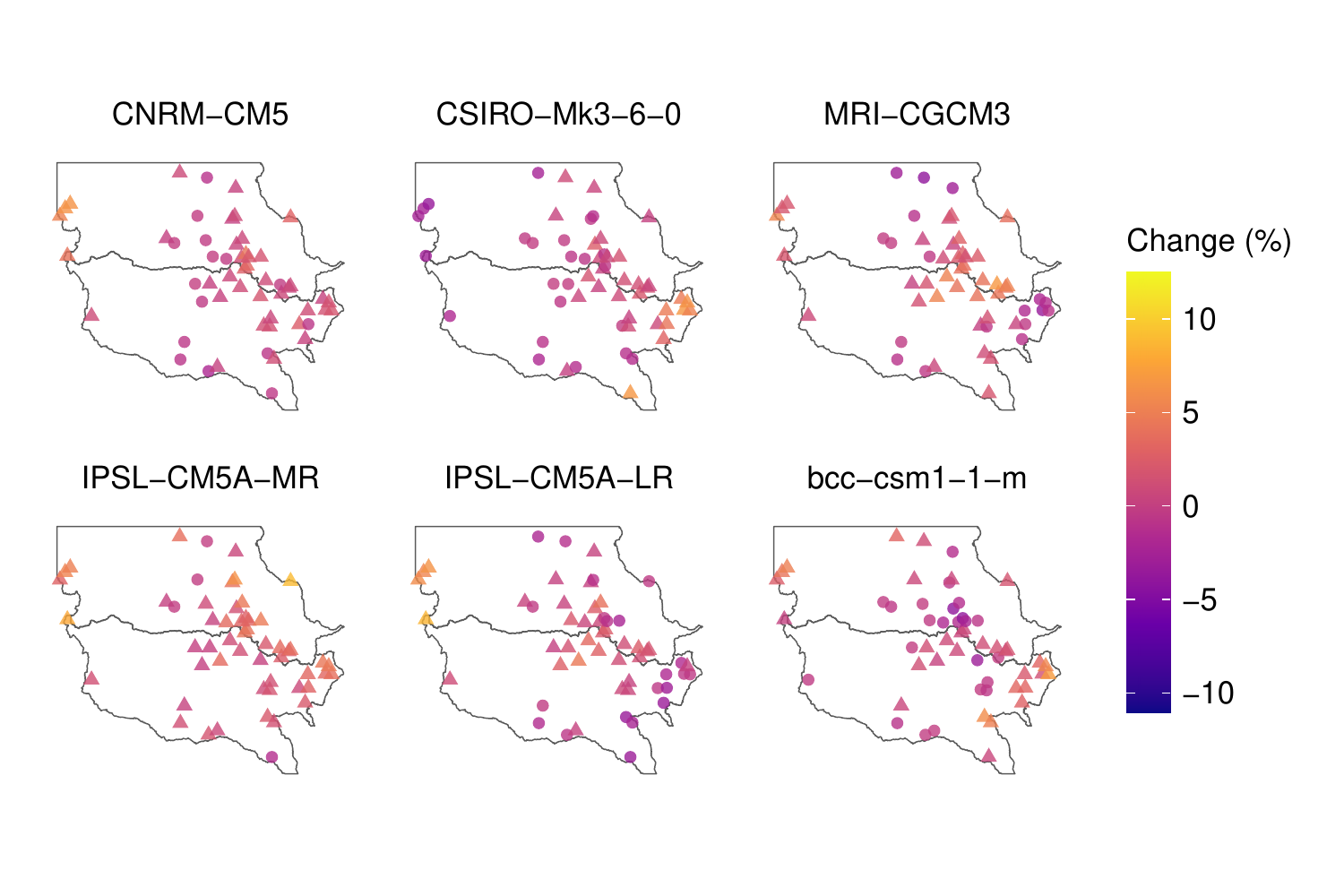}
\caption{\small Change in projected streamflow based on RCP 4.5.}
    \label{fig:rcp45project90}
    \end{subfigure}
    \hfill
    \begin{subfigure}{\linewidth}
    \includegraphics[scale=0.8]{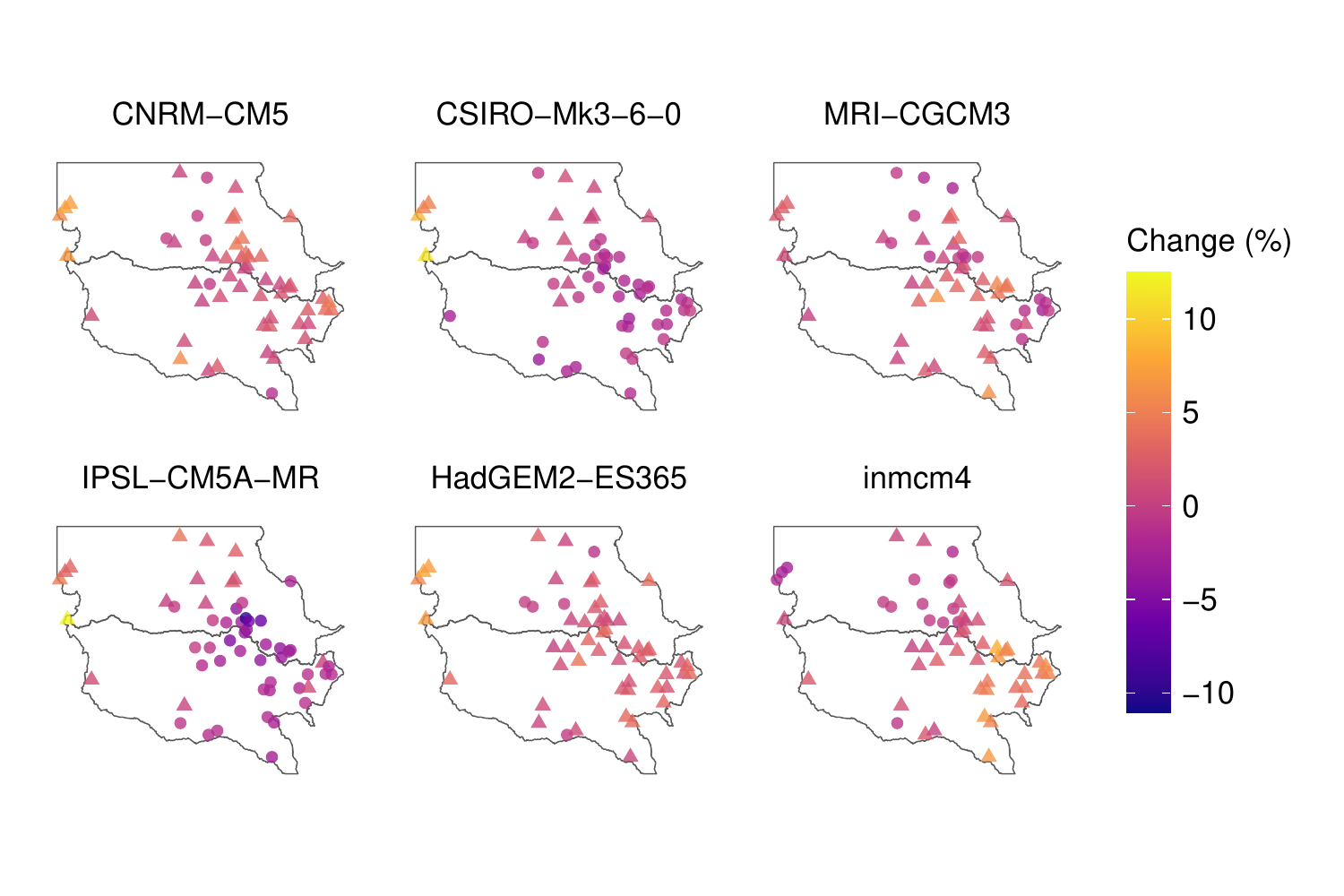}
\caption{\small Change in projected streamflow based on RCP 8.5.}
    \label{fig:rcp85project90}
    \end{subfigure}
    \caption{\small Percentage change in observed 0.90 quantile under RCP~4.5 and RCP~8.5 for 2006--2035, compared to the baseline period of 1972--2005. Triangles denote positive values and circles denote negative values.}
    \label{fig:q90}
\end{figure}

\begin{figure}
    \centering
    \begin{subfigure}{\linewidth}
    \includegraphics[scale=0.8]{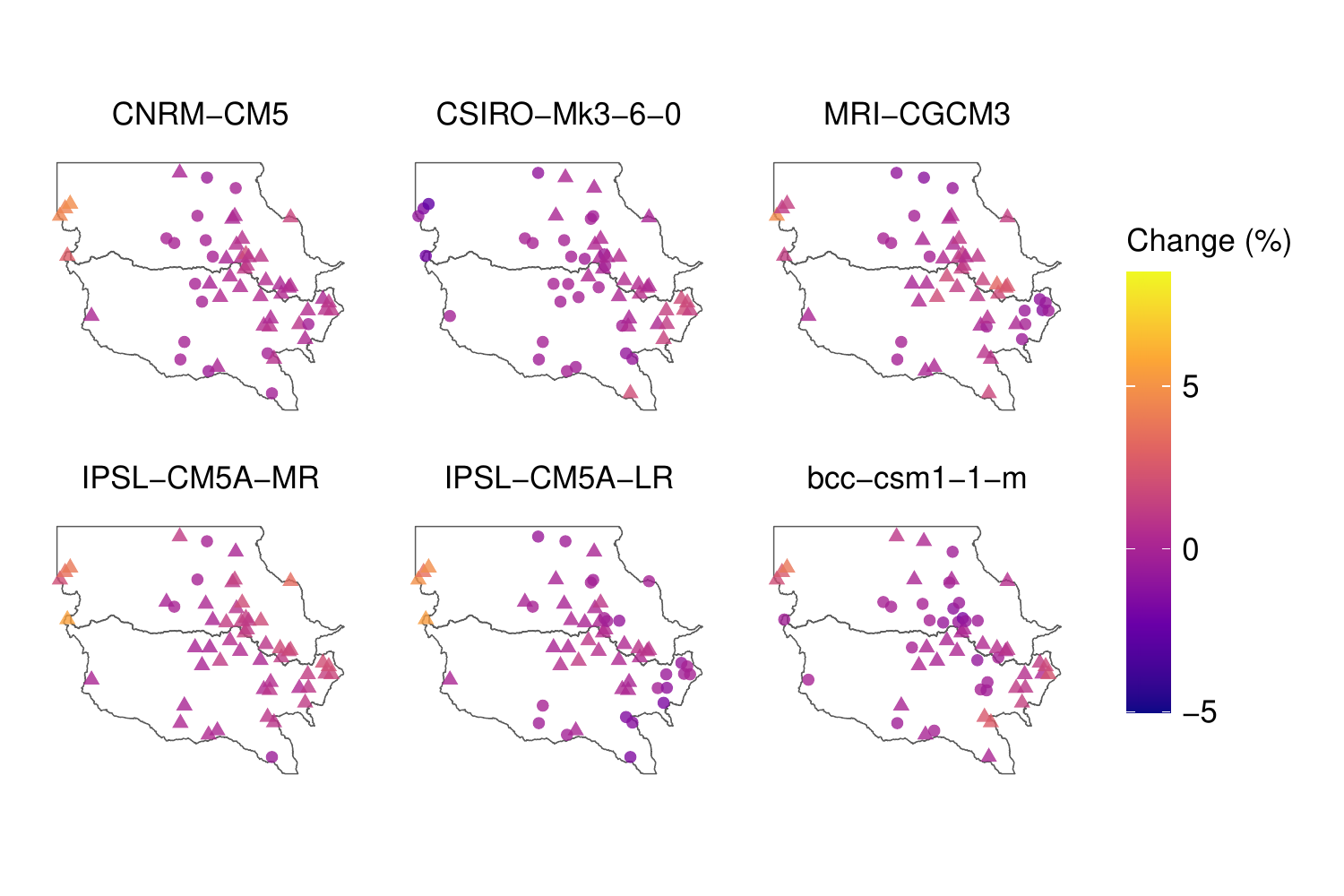}
\caption{\small Change in projected streamflow based on RCP 4.5.}
    \label{fig:rcp45project99}
    \end{subfigure}
    \hfill
    \begin{subfigure}{\linewidth}
    \includegraphics[scale=0.8]{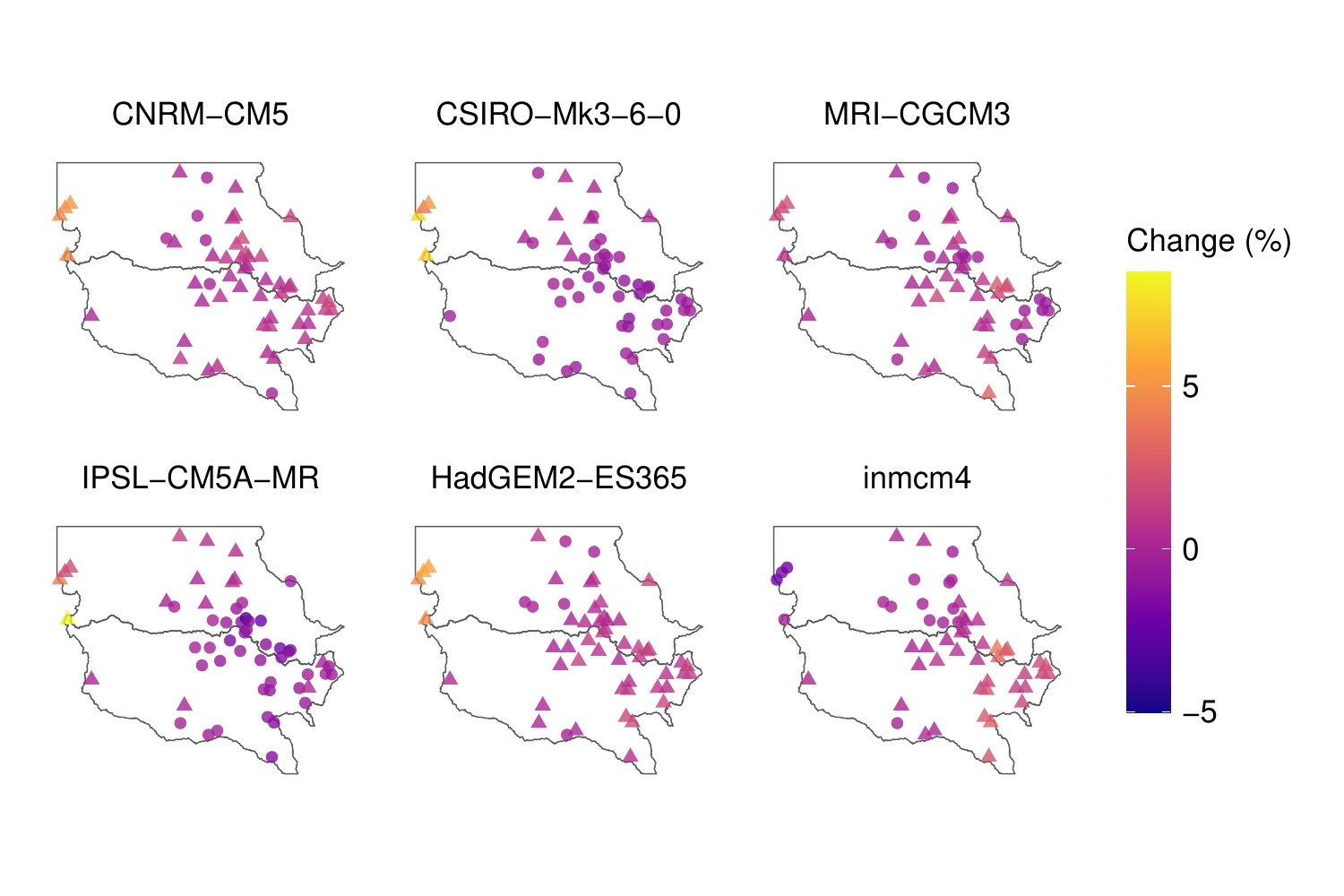}
\caption{\small Change in projected streamflow based on RCP 8.5.}
    \label{fig:rcp85project99}
    \end{subfigure}
    \caption{\small Percentage change in observed 0.99 quantile under RCP~4.5 and RCP~8.5 for 2006--2035, compared to the baseline period of 1972--2005. Triangles denote positive values and circles denote negative values.}
    \label{fig:q99}
\end{figure}
We used the bias-corrected MACA precipitation data for six RCP~4.5 and six RCP~8.5 models as specified in Section \ref{s:maca} to get future projections of streamflow. Future projections for MACA (and CMIP5 data in general) begin from 2005, and we consider the distribution of extremal streamflow forecasts for the period from 2006--2035. Each CMIP5 model also provides historical runs alongside the projections, from which we estimate the distribution of extremal streamflow for 1972--2005. For each scenario (historical, RCP 4.5, RCP 8.5) and each GCM model listed in Section \ref{s:maca}, we use seasonal and annual bias-corrected GCM precipitation to generate estimates of annual streamflow maxima using the following steps:
\begin{enumerate}
    \item Draw 1000 post burn-in samples $\btheta_1^{(1)},\ldots, \btheta_1^{(1000)}$ from the posterior distribution of the GEV parameters. Repeat steps 2--3 for each sample and each scenario
    \item Use bias-corrected GCM precipitation as covariates in \eqref{e:marginal} to get GEV distribution location, scale, and shape parameter estimates independently for each site
    \item Solve for and compute the 0.90 and 0.99 quantiles of the distribution of streamflow maxima over the entire time period.
\end{enumerate}
The quantiles for each site, given the GEV parameters for the entire time period (34 years for the historical period and 30 years for the projection period), can be computed by univariate root-finding algorithms. This gives us 1000 extremal quantile estimates of the distribution of annual streamflow maxima at each of the 55 sites for the historical, RCP 4.5, and RCP 8.5 scenarios. For each of the two RCP scenarios and two extremal quantile levels, we study and report the percent change in annual streamflow maxima compared to the historical period. 

Figures \ref{fig:q90}--\ref{fig:q99} show the mean percentage change in the observed $0.90$ and $0.99$ quantiles under the RCP 4.5 and RCP 8.5 projections, averaged over the 1000 estimates. The top row of each figure consists of models that project a wetter future, whereas the bottom row consists of models which project a drier future. In both figures, the triangles denote an increase, while the circles denote a decrease in annual streamflow maxima at each location. Four of the six models under each RCP scenario are common to both scenarios - CNRM-CM5, CSIRO-Mk3-6-0, and MRI-CGCM3 which project wetter futures, and IPSL-CM5A-MR, which projects a drier future. The output based on these four models can thus be compared across scenarios and quantile levels. For a particular quantile level, with the exception of CSIRO-Mk3-6-0, the wetter models predict more positive changes under RCP~8.5 than under RCP~4.5 Similarly, IPSL-CM5A-MR predicts more negative changes under RCP~8.5 than under RCP~4.5. CSIRO-Mk3-6-0 shows noticeable differences between RCP~4.5 and RCP~8.5 with several locations that show positive change under one scenario showing negative change under the other and vice versa.  We expect further divergences between scenarios if this study is extended to a longer time horizon due on the underlying assumptions of the 2 RCP scenarios.

Looking across quantile levels, we note that the 0.99 quantiles in Figure~\ref{fig:q99} estimate lower levels of change, ranging from -2.7\%--8.4\%, compared to the 0.90 quantiles in Figure \ref{fig:q90} which show changes between -10.3\%--12.3\%. However, the number of locations with positive changes are the same or higher when we go from the 0.90 quantile to the 0.99 quantile under both RCP scenarios. Under RCP~4.5, all six models estimate that more than 50\% locations have increased flow for both quantile levels, with values ranging from 51\% -- 93\%. For RCP~8.5, four out of the six models estimate more than half the locations to have increased streamflow. In this case, the values range from 22\% -- 91\%; in all cases, CSIRO-Mk3-6-0 gives the lowest estimates.

\begin{table}
\centering
\caption{\small {\bf Measure of joint exceedance in projected streamflow maxima}: Mean number of locations jointly above the 0.90 and 0.99 quantile thresholds. Values in parentheses represent the minimum and maximum projections from among the 6 models used in each scenario.}
\label{t:exceedance}
\begin{tabular}{ccccc}
 & \multicolumn{2}{c}{$u = 0.90$} & \multicolumn{2}{c}{$u=0.99$} \\\toprule
 & \textbf{RCP 4.5} & \textbf{RCP 8.5} & \textbf{RCP 4.5} & \textbf{RCP 8.5} \\\midrule
1972--2005 & (5.49,5.55) & (5.49,5.55) & (0.54,0.56)  & (0.54,0.56)  \\
2006--2035 & (5.49,5.54) & (5.50,5.53) & (0.54,0.56) & (0.55,0.56)\\\bottomrule
\end{tabular}
\end{table} 

Table \ref{t:exceedance} shows the expected number of locations jointly above the threshold for the historical and projection periods based on Monte-Carlo simulations from the fitted spatial model using bias-corrected GCM precipitation data. The values in parentheses correspond to the minimum and maximum of the estimates obtained from the 6 GCM models used. If the probability of exceeding the threshold at all locations were independent, the number of locations above the threshold would follow a Binomial distribution with parameters $n=55$ and probability $1-u$, for the two cases of $u=0.90$ and $u=0.99$. In turn, the expected number of locations above the threshold under the independence assumption would be $5.5$ and $0.55$ respectively. For both the historical and projection periods, estimates from most of the models are higher than estimates from the independence assumption. In particular, both the mean and median for each of the 8 sets of values are higher than what we would get from an independence assumption. Overall, this suggests that concurrent extremal streamflow at multiple locations is likely to keep occurring into the near future.

\section{Discussion}
\label{}
In this paper, we propose a non-stationary process mixture model for spatial extreme value analysis. The marginal distributions of the process are GEV, while the spatial dependence is specified as an interpolation of a GP and an MSP indexed by a weight parameter which is allowed to vary spatio-temporally, introducing non-stationarity. Similarly, STVC specifications used for the marginal parameters make the model flexible in terms of learning different spatio-temporal patterns present in the data. The model is an extension of the (stationary) process mixture model introduced in \cite{MajumderReichShaby2022}. The intractable joint likelihood for the spatial model is approximated using a Vecchia decomposition, and is learned using the density regression approach of \cite{xu-reich-2021-biometrics}. The density regression estimates a quantile process for the approximate likelihood whose weights are obtained from a neural network by maximizing the approximate likelihood.

We use the NPMM to provide climate informed near-term projections of annual streamflow maxima for the central US region. The CUS is affected by convective storms and, therefore, any projections of streamflow should take into account seasonal and annual precipitation over the region. The CUS is divided into two HUC-02 regions, and the asymptotic regime for the regions are estimated independently. We used observed NClimGrid precipitation data to fit the model for annual streamflow maxima. The means of the posterior distribution puts Region 11 in the south to be asymptotically dependent for all 50 years, whereas Region 10L in the north is asymptotically dependent for 39 out of the 50 years and asymptotically independent for the rest of the years. Region 10L also has more variability in the posterior mean of asymptotic (in)dependence parameter from year to year. These inter-year differences and differences between the regions justify the appropriateness of the non-stationary assumptions we make about the process. While we find no significant linear relationship between region-wide precipitation and the logit of weight parameter, we note that region 11 has higher precipitation compared to region 10L. Afterwards, bias-corrected GCM precipitation projections are used as covariates to obtain streamflow estimates for the future period of 2006--2035 and compared against the historical period of 1972--2005. Based on our projections, both the magnitude of extremal streamflow as well as the number of locations which are concurrently affected by these extreme events are likely to increase in the near-term future.

Future research will focus on generating long-term climate-informed projections. The current work considers only seasonal precipitation as covariates, as adding too many variables adversely affected MCMC convergence. However, longer-term precipitation as well as temperature can affect streamflow 
 \cite{Awasthietal2022}, and we would like to incorporate additional covariates in future work. Learning the weight parameter proves more challenging for the NPMM compared to its stationary equivalent; we hope to improve the spatial dependence in the model as well as the estimates obtained from it by incorporating network structure, as has been done for both max-stable \citep{asadi2015extremes} and Gaussian \citep{santos2022bayesian} processes. Relaxing the simplifying assumptions on the smoothness and range parameters would improve the spatial modeling, but could make estimation more difficult as more variables are free to vary. Finally, the synthetic likelihood approach to density estimation for spatial processes using deep learning is not specific to the NPMM, and we would like to explore its performance and properties for other spatial extremes models.

\section*{Acknowledgments}
The authors thank Prof. Sankarasubramanian Arumugam of NC State University for discussion of the data and scope of the project.
\section*{Funding}
This work was supported by grants from the Southeast National Synthesis Wildfire and the United States Geological Survey's National Climate Adaptation Science Center (G21AC10045) and the National Science Foundation (DMS2152887, CBET2151651). Part of this research was performed while author Reetam Majumder was visiting the Institute for Mathematical and Statistical Innovation (IMSI), which is supported by the National Science Foundation (Grant No. DMS-1929348).

\appendix
\section{Derivation of Conditional Exceedance for a Common Spatial Process}\label{s:chi_derivation}
\citep{MajumderReichShaby2022} derived $\chi(\bs_1,\bs_2)$ for a process mixture model with a common MSP $R(\bs_1) = R(\bs_2) = R$ and $W(\bs_1)$ and $W(\bs_2)$ are independent. We extend that and focus on a specific case where $\delta_1 = \delta$ and $\delta_2 = 1-\delta$, where $\delta_1, \delta_2$ are defined as in Section \ref{s:tail_behavior}. This is a convenient case because with this restriction both sites have the same marginal distribution. This case is also interesting because it illustrates the behavior of the process when the two sites are in different asymptotic regimes. We denote $g_W\{W(\bs_1)\} = W_1^*$, $g_W\{W(\bs_2)\} = W_2^*$, $g_R(R) = R^*$ for convenience. By assumption $W_1^*,W_2^*,R^*\iid \mbox{Exponential}(1)$. Under these conditions, the joint survival probability is as follows:
\begin{align*}
    Pr[Y_1>y,Y_2>y] &= Pr[\delta_1R^* + (1-\delta_1)W_1^*>y, \delta_2R^* + (1-\delta_2)W_2^*>y]\\
    &= \mathbb E_{R^*}\biggl[Pr\bigl\{W_1^*>\frac{y-\delta r}{1-\delta}\bigl\}Pr\bigl\{W_1^*>\frac{y-(1-\delta) r}{\delta}\bigl\}|R^*=r \biggr].
\end{align*}
Defining $r_1 := (y-\delta r)/(1-\delta)$ and $r_2 := (y-(1-\delta) r)/\delta$, we get
\begin{multline}\label{e:chi_num}
    Pr[Y_1>y,Y_2>y] = \mathbb E_{R^*} \biggl[Pr\{W_1^*>r_1\}Pr\{W_2^*>r_2\}\mathbb I\{r_1>0,r_2>0\}\biggr]\\
    + \mathbb E_{R^*} \biggl[Pr\{W_1^*>r_1\}\mathbb I\{r_1>0,r_2<0\}\biggr]\\
    + \mathbb E_{R^*} \biggl[Pr\{W_2^*>r_2\}\mathbb I\{r_1<0,r_2>0\}\biggr] + \mathbb E_{R^*}\biggl[ \mathbb I\{r_1<0,r_2<0\}\biggr].
\end{multline}
Note that:
\begin{align*}
    r_1>0, r_2>0 &\implies r < \min(y/\delta,y/(1-\delta))\\
    r_1>0, r_2<0 &\implies (y/(1-\delta) < r < y/\delta)\mathbb I\{\delta<0.5\}\\
    r_1<0, r_2>0 &\implies (y/\delta < r < y/(1-\delta))\mathbb I\{\delta>0.5\}\\
    r_1<0, r_2<0 &\implies r>\max(y/\delta,y/(1-\delta))
\end{align*}

We first assume that $\delta<0.5$. Denoting the four terms on the right-hand side of \eqref{e:chi_num} as $J_1,J_2,J_3$, and $J_4$, we first see that $J_3 = 0$. The remaining three terms are computed individually.
\begin{align*}
    J_1 &= \exp\bigl\{-y\bigl(\frac{1}{\delta}+ \frac{1}{1-\delta}\bigr)\bigr\}\int_0^{y/1-\delta}\exp\bigl\{r\bigl(\frac{\delta}{1-\delta} + \frac{1-\delta}{\delta}\bigr)\bigr\}\exp\{-r\} dr\\
    &= \exp\bigl\{-y\bigl(\frac{1}{\delta}+ \frac{1}{1-\delta}\bigr)\bigr\}\int_0^{y/1-\delta}\exp\bigl\{\frac{3\delta^2 - 3\delta + 1}{\delta(1-\delta)}r\bigr\} dr\\
        &= k_1 \exp\bigl\{-y\bigl(\frac{1}{\delta}+ \frac{1}{1-\delta}\bigr)\bigr\}\bigl[\exp\bigl\{\frac{3\delta^2 - 3\delta + 1}{\delta(1-\delta)^2}y\bigr\} -1\bigr]\\
    &= k_1 \exp\bigl\{-\frac{y}{1-\delta}\bigr\}\bigl[\exp\bigl\{-\frac{1-2\delta}{(1-\delta)^2}y\bigr\} - \exp\bigl\{-\frac{y}{\delta}\bigr\} \bigr],
\end{align*}
where $k_1$ is the appropriate constant arising from the integration.
\begin{align*}
    J_2 &= \exp\bigl\{-\frac{y}{1-\delta}\bigr\}\int_{y/1-\delta}^{y/\delta}\exp\bigl\{r\bigl(\frac{\delta}{1-\delta}-1\bigr)\bigr\}dr\\
    &= k_2 \exp\bigl\{-\frac{y}{1-\delta}\bigr\}\bigl[\exp\bigl\{-\frac{1-2\delta}{\delta(1-\delta)}y\bigr\} - \exp\bigl\{-\frac{1-2\delta}{(1-\delta)^2}y\bigr\}\bigr],
\end{align*}
where $k_2$ is the appropriate constant that arises from the integration. Finally,
\begin{align*}
    J_4 &= \exp\{-y/\delta\}.
\end{align*}
The marginal survival probability can be obtained from \eqref{eq:hypo}. We denote it as $M$, where
\begin{align*}
    M = \frac{\delta}{1-2\delta}\exp\{-\frac{y}{\delta}\} - \frac{1-\delta}{1-2\delta}\exp\{-\frac{y}{1-\delta}\}.
\end{align*}
The conditional exceedance probability $\chi(\bs_1,\bs_2)$ can be expressed as:
\begin{align*}
    \chi(\bs_1,\bs_2) &= \lim_{y\to\infty}\frac{J_1+J_2+J_3+J_4}{M}\\
    &= \lim_{y\to\infty}\frac{J_1}{M} + \lim_{y\to\infty}\frac{J_2}{M} + \lim_{y\to\infty}\frac{J_4}{M}.
\end{align*}
Each of the limits are evaluated individually:
\begin{align*}
    \frac{J_1}{M} &= k_1 \frac{\exp\bigl\{-\frac{1-2\delta}{(1-\delta)^2}y\bigr\} - \exp\bigl\{-\frac{y}{\delta}\bigr\}}{\frac{\delta}{1-2\delta}\exp\bigl\{-y\frac{1-2\delta}{\delta(1-\delta)}\bigr\} - \frac{1-\delta}{1-2\delta} }\\
    \implies \lim_{y\to \infty}\frac{J_1}{M} &= k_1\frac{0-0}{0 - \frac{1-\delta}{1-2\delta}} = 0.
\end{align*}
\begin{align*}
    \frac{J_2}{M} &= k_2 \frac{\exp\bigl\{-\frac{1-2\delta}{\delta(1-\delta)}y\bigr\} - \exp\bigl\{-\frac{1-2\delta}{(1-\delta)^2}y\bigr\}}{\frac{\delta}{1-2\delta}\exp\bigl\{-y\frac{1-2\delta}{\delta(1-\delta)}\bigr\} - \frac{1-\delta}{1-2\delta}}\\
    \implies \lim_{y\to \infty}\frac{J_2}{M} &= 0
\end{align*}
Finally,
\begin{align*}
    \frac{J_4}{M} &= \frac{\exp\bigl\{-\frac{y}{\delta}\bigr\}}{\frac{\delta}{1-2\delta}\exp\{-\frac{y}{\delta}\} - \frac{1-\delta}{1-2\delta}\exp\{-\frac{y}{1-\delta}\}}\\
    &= \frac{ \exp\bigl\{ -y\frac{1-2\delta}{\delta(1-\delta)} \bigr\}   }{ \frac{\delta}{1-2\delta}\exp\bigl\{-y\frac{1-2\delta}{\delta(1-\delta)}\bigr\} - \frac{1-\delta}{1-2\delta} }\\
    \implies \lim_{y\to \infty} \frac{J_4}{M} &= 0.
\end{align*}
$$\therefore \chi(\bs_1,\bs_2) = 0.$$
Next, consider the case of $\delta > 0.5$. We see that the term $J_2$ in \eqref{e:chi_num} is 0. Like before, we simplify the remaining 3 terms.
\begin{align*}
 J_1 &= \exp\bigl\{-y\bigl(\frac{1}{\delta}+ \frac{1}{1-\delta}\bigr)\bigr\}\int_0^{y/\delta}\exp\bigl\{r\bigl(\frac{\delta}{1-\delta} + \frac{1-\delta}{\delta}\bigr)\bigr\}\exp\{-r\} dr\\
    &= \exp\bigl\{-y\bigl(\frac{1}{\delta}+ \frac{1}{1-\delta}\bigr)\bigr\}\int_0^{y/\delta}\exp\bigl\{\frac{3\delta^2 - 3\delta + 1}{\delta(1-\delta)}r\bigr\} dr\\
        &= k_3 \exp\bigl\{-y\bigl(\frac{1}{\delta}+ \frac{1}{1-\delta}\bigr)\bigr\}\bigl[\exp\bigl\{\frac{3\delta^2 - 3\delta + 1}{\delta^2(1-\delta)}y\bigr\} -1\bigr]\\
    &= k_3 \exp\bigl\{-\frac{y}{\delta}\bigr\}\bigl[\exp\bigl\{-\frac{2\delta-1}{\delta^2}y\bigr\} - \exp\bigl\{-\frac{y}{1-\delta}\bigr\} \bigr],
\end{align*}
where $k_3$ is the appropriate constant from the integration. We note the symmetry between $J_1$ for $\delta<0.5$ and $J_1$ computed for $\delta>0.5$. It is straightforward to show that $\lim_{y\to \infty} J_1/M = 0$ in this case as well. It follows by symmetry that $\lim_{y\to \infty} J_4/M = 0$ for $\delta>0.5$. Finally, we verify the behavior for $J_3$:
\begin{align*}
     J_3 &= \exp\bigl\{-\frac{y}{\delta}\bigr\}\int_{y/\delta}^{y/1-\delta}\exp\bigl\{r\bigl(\frac{1-\delta}{\delta}-1\bigr)\bigr\}dr\\
    &= k_4 \exp\bigl\{-\frac{y}{\delta}\bigr\}\bigl[\exp\bigl\{-\frac{2\delta-1}{\delta(1-\delta)}y\bigr\} - \exp\bigl\{-\frac{2\delta-1}{\delta^2}y\bigr\}\bigr],
\end{align*}
where $k_4$ is the appropriate constant for integration. Thus, $\lim_{y\to \infty}J_3/M = 0$ due to its symmetry with $J_2$. 

Therefore, for $\delta \in (0,0.5) \cup (0.5,1)$, $\chi(\bs_1,\bs_2) = 0$.

\section{Computational Details}
\subsection{Asymptotic joint tail behavior} 
\begin{figure}
    \centering
    \begin{subfigure}{0.49\linewidth}
    \includegraphics[width=\linewidth]{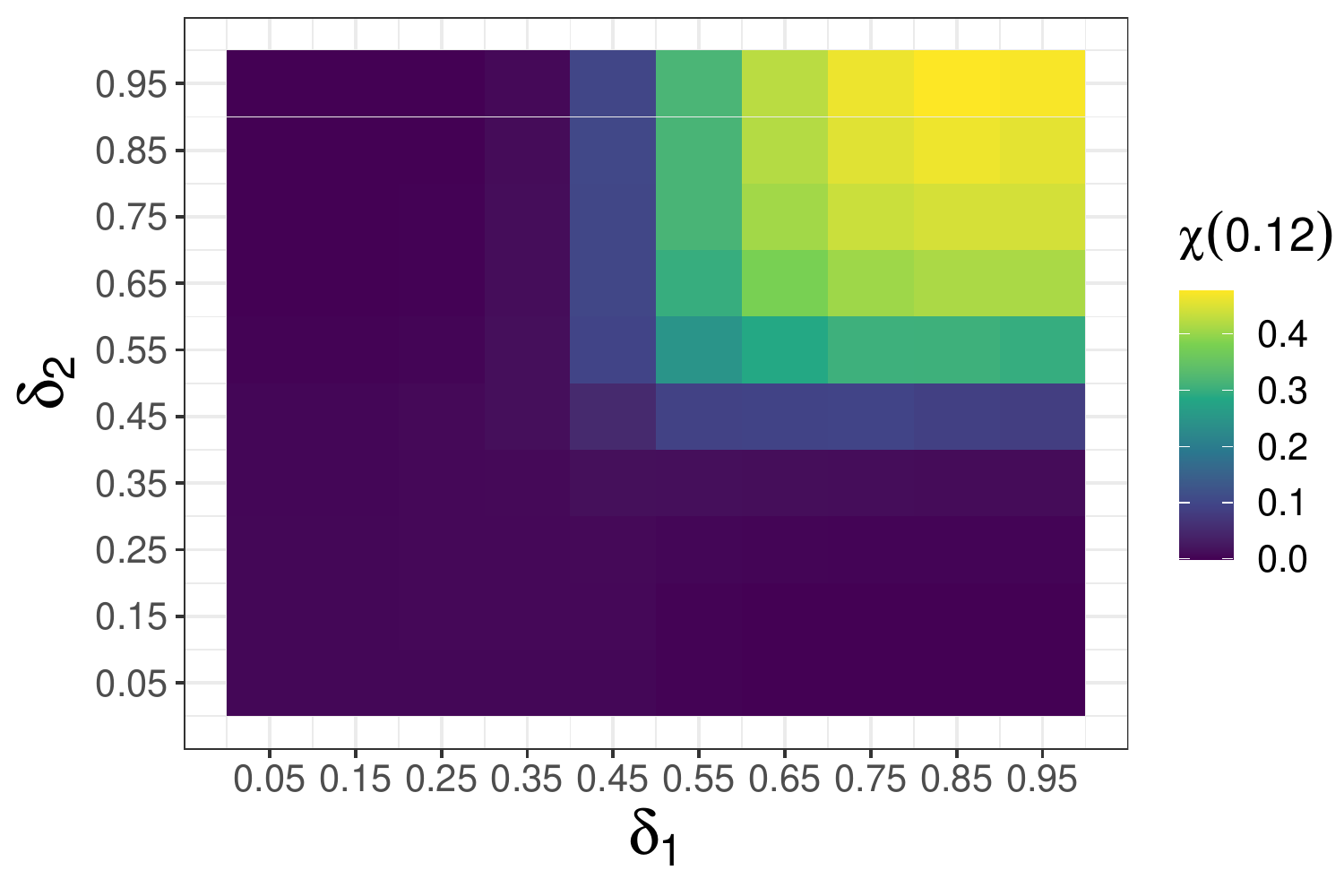}
\caption{\small $\chi_u(h)$ when $\rho_R = \rho_W$.}
    \label{fig:chi_sq_rho}
    \end{subfigure}
    \hfill
    \begin{subfigure}{0.49\linewidth}
    \includegraphics[width=\linewidth]{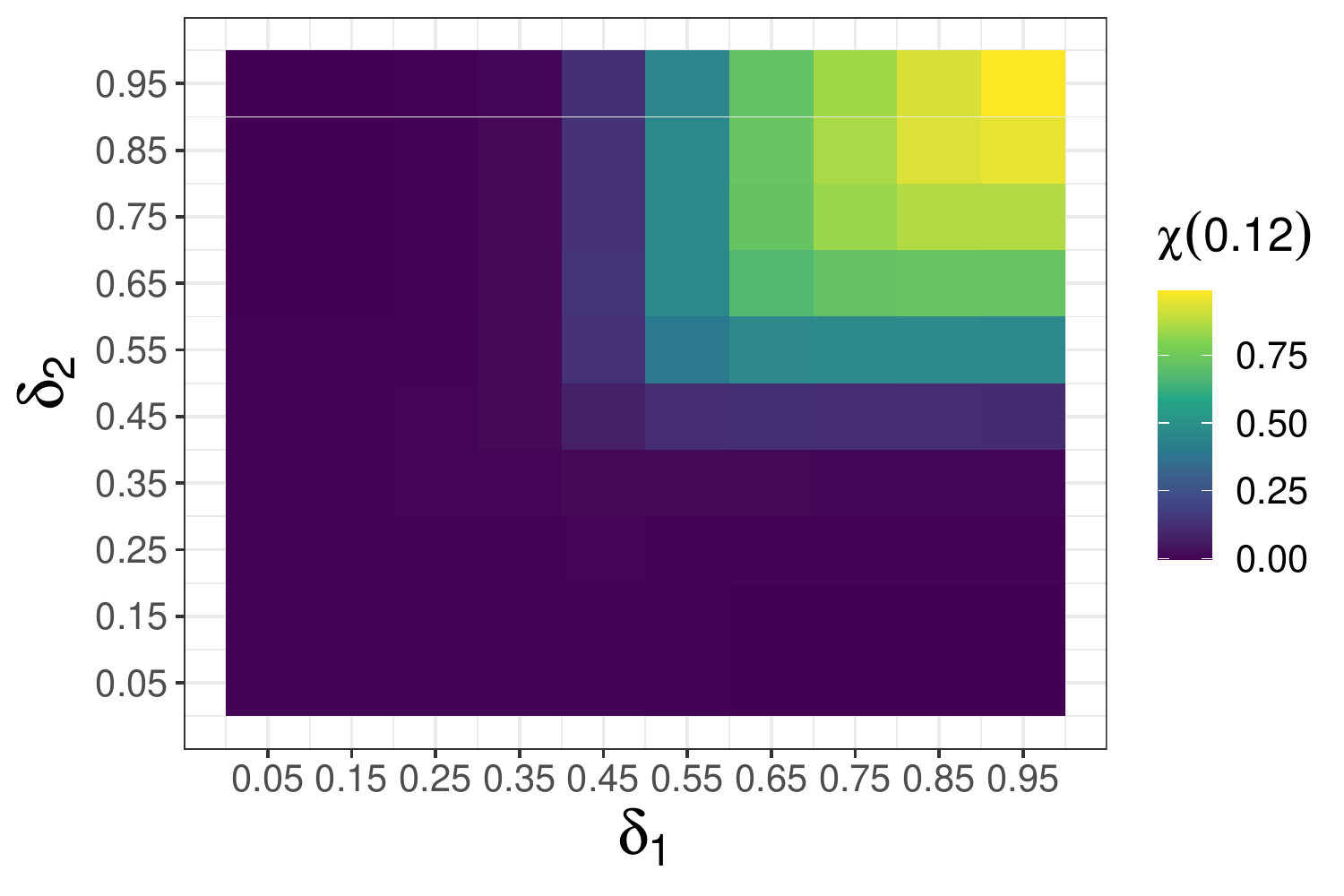}
\caption{\small $\chi_u(h)$ when $R_t(\bs)=R$.}
    \label{fig:chi_sq_HW}
    \end{subfigure}
    \caption{\small Empirical $\chi_u(h)$ for different combinations of $\delta_1$ and $\delta_2$ with threshold $u=0.9999$ under two different model specifications.}
    \label{fig:empericalchi_2}
\end{figure}

Figure \ref{fig:empericalchi_2} depicts the behavior of $\chi_u(0.12)$ at the 0.9999 quantile for two related models, which relax our current model assumption of $\rho_R = 0.19\rho_W$. In Figure \ref{fig:chi_sq_rho}, we assume that $\rho_R = \rho_W$. This increases the range of $\chi_u(0.12)$ as more extremal dependence is introduced. In Figure \ref{fig:chi_sq_HW}, we replace the MSP with a GEV(1,1,1) distribution, which makes this equivalent to the model presented in \cite{Huser-Wadsworth}. This has the maximum amount of extremal dependence among this class of models by construction, which is reflected in the high range of $\chi_u(0.12)$. However, for both cases, the same behavior holds for different values of $\delta_1$ and $\delta_2$, with asymptotic dependence only if both sites are in an asymptotic dependence regime.

\subsection{Variable importance plots}\label{s:VI-plots}
\begin{figure}
    \centering
    \begin{subfigure}{0.49\linewidth}
    \includegraphics[width=\linewidth]{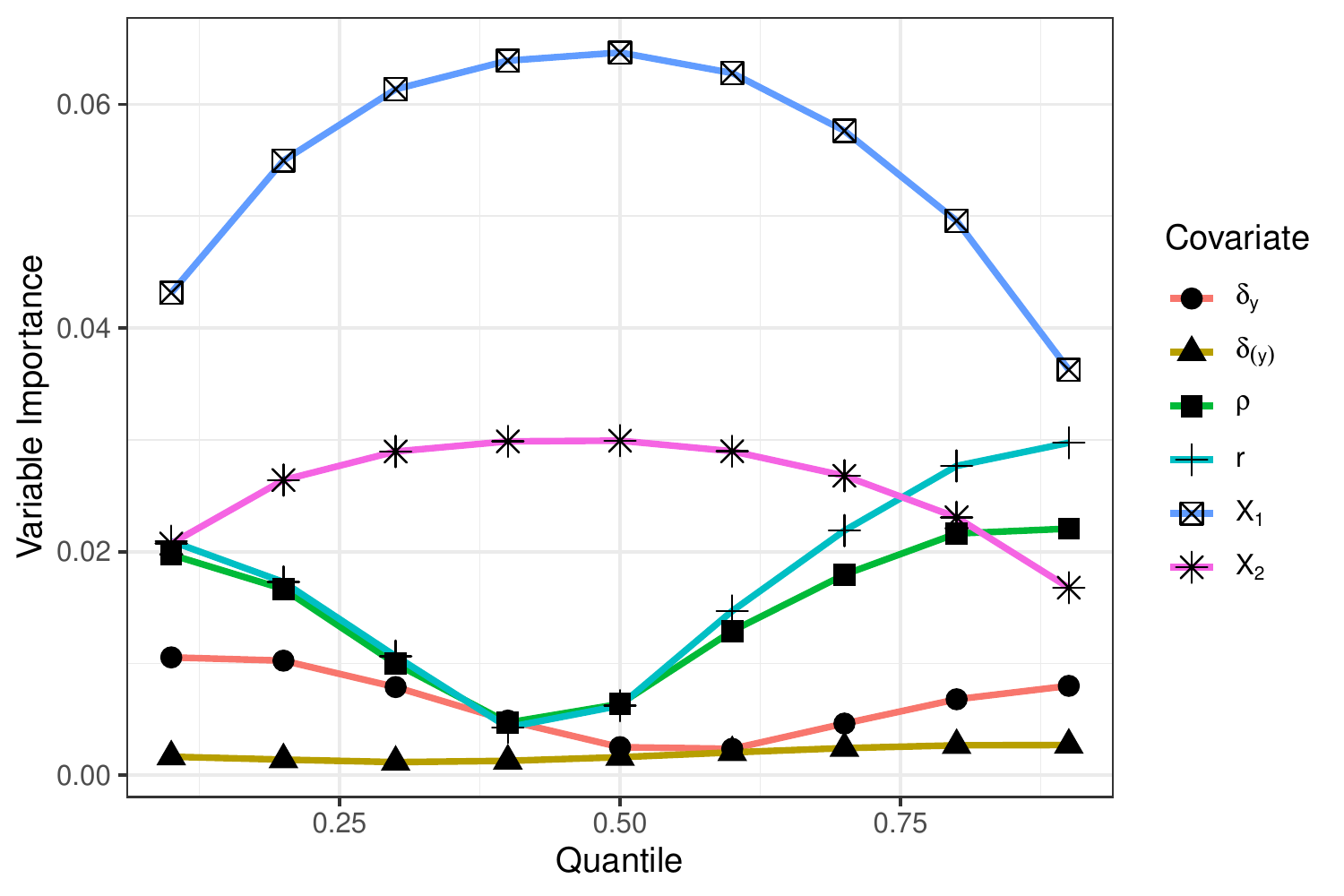}
\caption{\small VI for location 11.}
    \label{fig:VI_11}
    \end{subfigure}
    \hfill
    \begin{subfigure}{0.49\linewidth}
    \includegraphics[width=\linewidth]{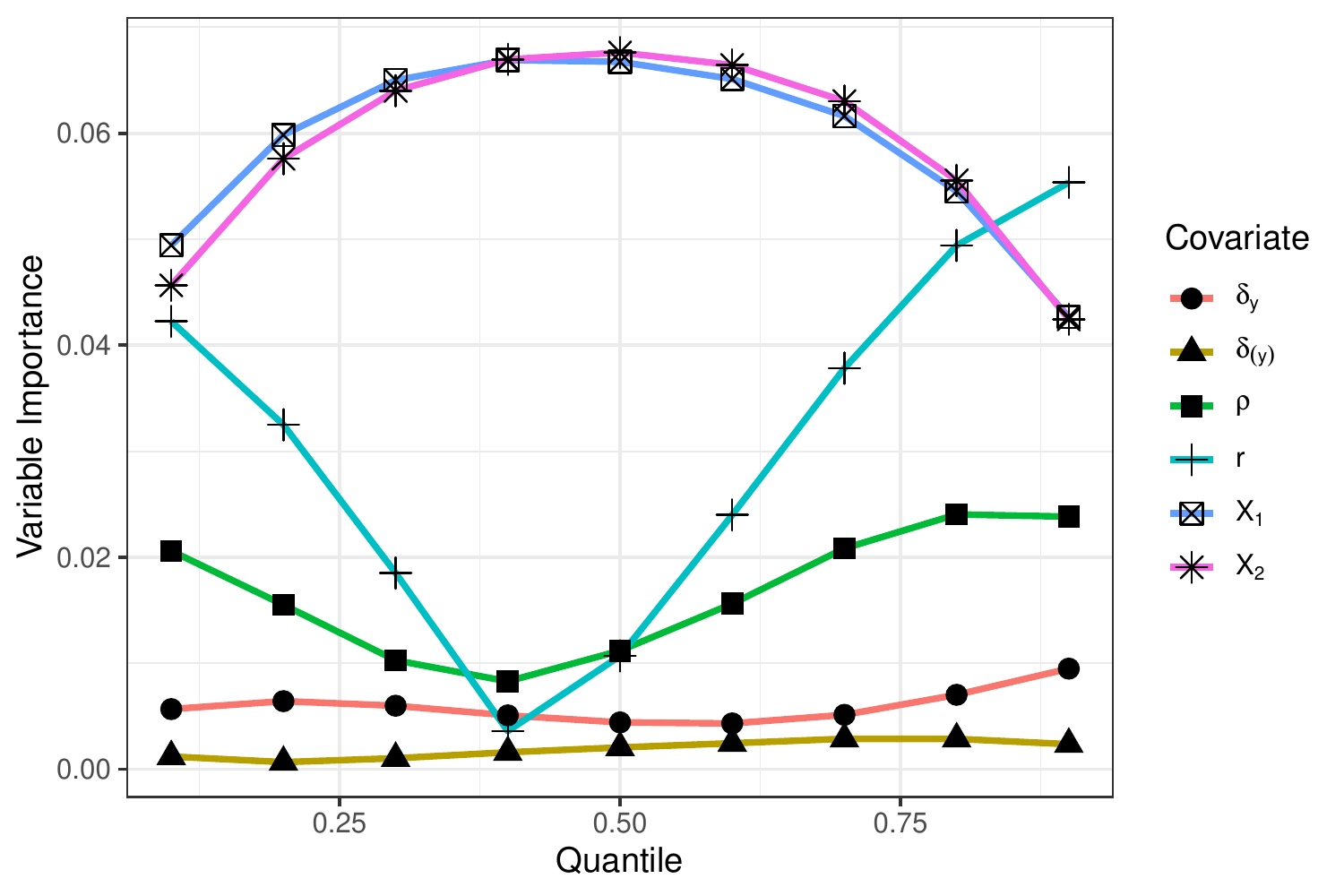}
\caption{\small VI for location 16.}
    \label{fig:VI_16}
    \end{subfigure}
    \\[3ex]
     \begin{subfigure}{0.49\linewidth}
    \includegraphics[width=\linewidth]{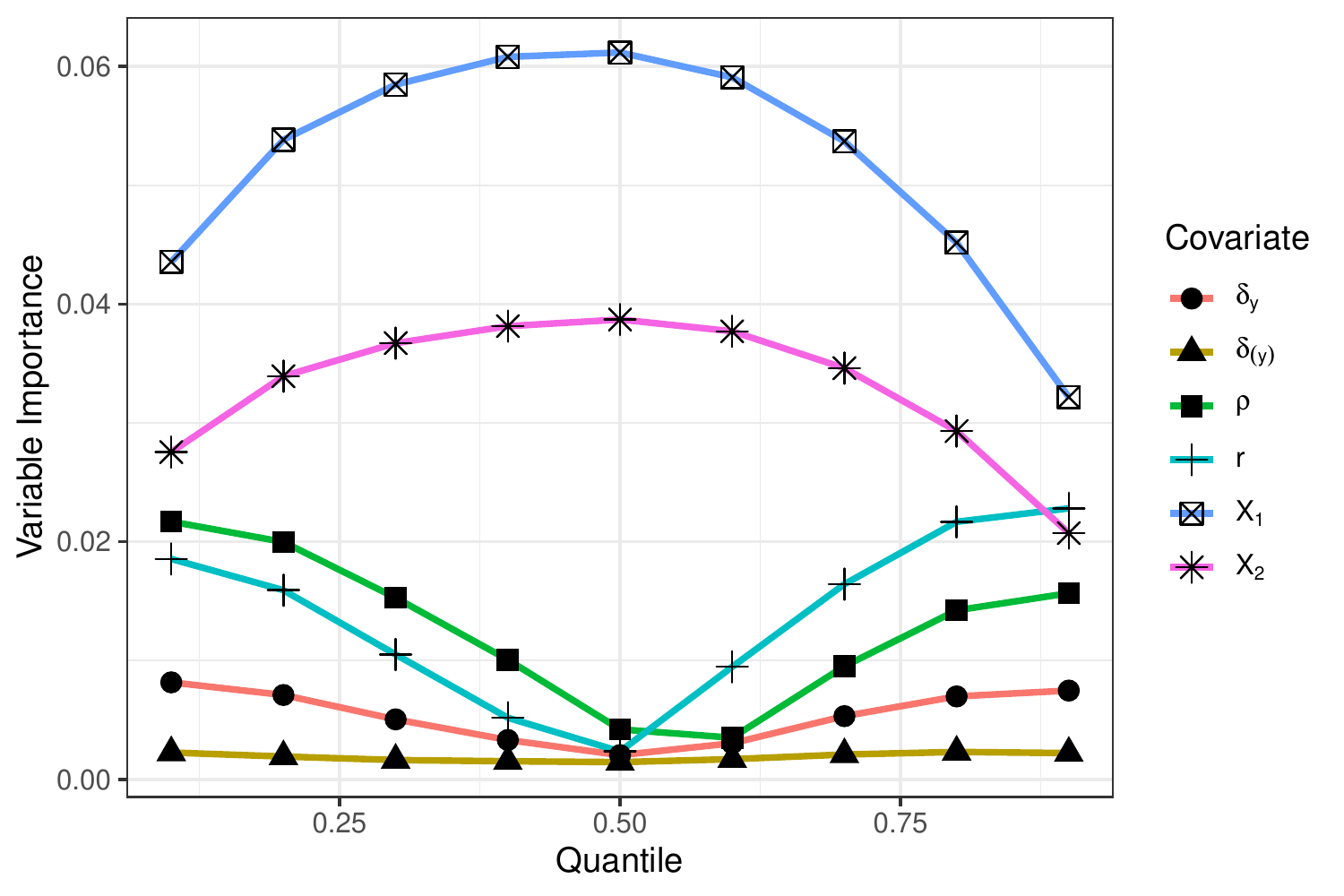}
\caption{\small VI for location 35.}
    \label{fig:VI_35}
    \end{subfigure}
    \hfill
     \begin{subfigure}{0.49\linewidth}
    \includegraphics[width=\linewidth]{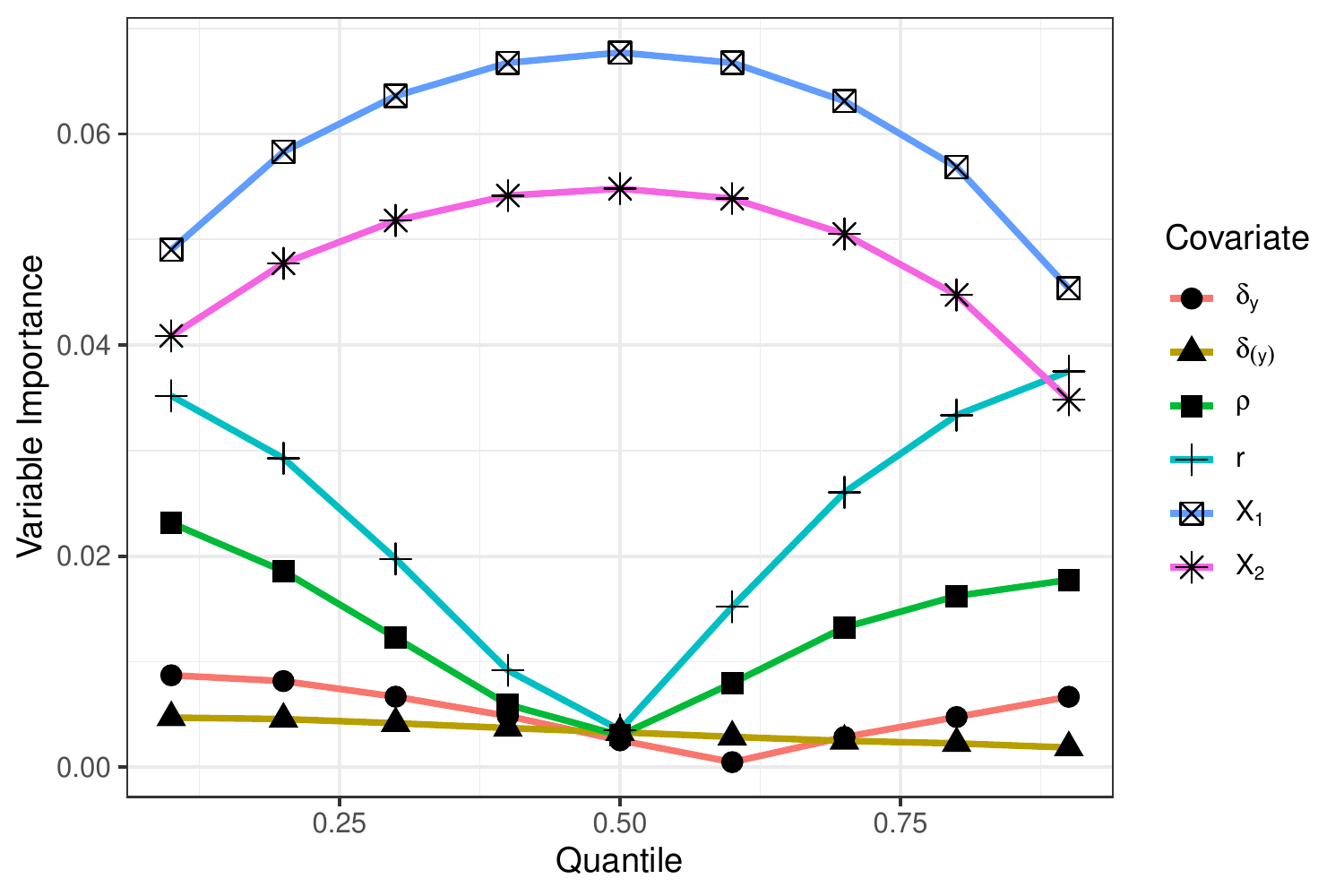}
\caption{\small VI for location 50.}
    \label{fig:VI_50}
    \end{subfigure}
    \caption{\small Variable importance (VI) plots based on SPQR output for 4 different locations within the CUS.}
    \label{fig:VI_2}
\end{figure}

Figure \ref{fig:VI_2} presents variable importance plots for 4 different locations within our study area. Location 11 does not have a full suite of neighbors, as the Vecchia neighboring set can have up to 15 neighbors. Location 16 is the first location which has all 15 neighbors, and locations 35 and 50 also have all 15 neighbors. For all 4 locations, the nearest neighbor has the highest importance. The importance of the second neighbor varies from location to location. We have found this to be a function of the spatial configuration - in particular, how far the second neighbor is from the response site, as well as how close it is to the other neighbors. It could also depend on whether it belongs to the same region or not. 

The remainder of the neighbors show similar behavior with a steady drop off of their importances, and have thus been omitted for clarity. It is interesting to note the fundamentally different way the neighbors affect the quantiles of the response compared to how the spatial parameters affect them. The neighbors have the largest effect around the median and drop off in importance near the extreme quantiles at both ends. The spatial parameters have the opposite behavior. We also note that $\delta_y$ is more important to the response compared to $\delta_{(y)}$. This is to be expected since $\delta_y$ is the mixing parameter that corresponds to the response, while $\delta_{(y)}$ can be either 0 or a function of the other mixing parameter that does not directly affect the response.

\subsection{Parameter estimates}\label{s:marg-estimates}

\begin{table}
\footnotesize
\centering
\caption{\footnotesize \textbf{Model fit diagnostics for marginal GEV parameters:} Standard errors based on the maximum likelihood estimates of GEV distributions fitted independently at each location (MLE), and posterior standard deviations based on the process mixture model (NPMM). Values represent an average taken over all 55 locations.}
\label{t:GEV_fits}
\begin{tabular}{ccc|ccc}
\toprule
\textbf{Parameter} & \textbf{MLE} & \textbf{NPMM} & \textbf{Parameter} & \textbf{MLE} & \textbf{NPMM} \\\midrule
$\mu_0$ & 0.05 & 0.03 & $\mu_1$ & 0.17 & 0.09 \\
$\mu_2$ & 0.07 & 0.05 & $\mu_3$ & 0.09 & 0.06 \\
$\mu_4$ & 0.08 & 0.05 & $\mu_5$ & 0.07 & 0.05 \\
$\sigma$ & 0.20 & 0.01 & $\xi$ & 0.22 & 0.13 \\\bottomrule
\end{tabular}
\end{table}

\begin{table}
\footnotesize
\centering
\caption{\footnotesize {\bf STVC parameter estimates}: Mean and SD for the GP parameters for the marginal GEV parameters.}
\label{t:SVC_params}
\begin{tabular}{ccc|ccc|ccc}
\toprule
\textbf{Param.} & \textbf{Mean} & \textbf{SD} & \textbf{Param.} & \textbf{Mean} & \textbf{SD} & \textbf{Param.} & \textbf{Mean} & \textbf{SD} \\\midrule
$\beta_{\mu_0}$ & -0.01 & 0.18 & $\tau^2_{\mu_0}$ & 0.19 & 0.04 & $\rho_{\mu_0}$ & 4.52 & 1.60 \\
$\beta_{\mu_1}$ & -0.06 & 0.25 & $\tau^2_{\mu_1}$ & 0.26 & 0.08 & $\rho_{\mu_1}$ & 3.24 & 1.53 \\
$\beta_{\mu_2}$ & 0.20 & 0.29 & $\tau^2_{\mu_2}$ & 0.30 & 0.09 & $\rho_{\mu_2}$ & 2.86 & 1.46 \\
$\beta_{\mu_3}$ & 0.26 & 0.32 & $\tau^2_{\mu_3}$ & 0.33 & 0.11 & $\rho_{\mu_3}$ & 2.56 & 1.43 \\
$\beta_{\mu_4}$ & 0.06 & 0.24 & $\tau^2_{\mu_4}$ & 0.25 & 0.07 & $\rho_{\mu_4}$ & 3.50 & 1.55 \\
$\beta_{\mu_5}$ & 0.08 & 0.22 & $\tau^2_{\mu_5}$ & 0.23 & 0.06 & $\rho_{\mu_5}$ & 3.73 & 1.59 \\
$\beta_{\sigma}$ & 0.17 & 2.19 & $\tau^2_{\sigma}$ & 0.89 & 0.48 & $\rho_{\sigma}$ & 1.36 & 1.20 \\
$\beta_{\xi}$ & 0.33 & 0.69 & $\tau^2_{\xi}$ & 0.72 & 0.29 & $\rho_{\xi}$ & 1.57 & 1.14\\\bottomrule
\end{tabular}
\end{table}

Table \ref{t:GEV_fits} provides a comparison of the marginal GEV model fits across locations based on the NPMM, as well as independent MLE estimates of the GEV parameters. The MLE estimates were used as initial values in our MCMC; we computed the standard errors for each variable and averaged it across the 55 sites. For the NPMM estimate, we compute the posterior SD of each parameter based on 20,000 post-burn in samples, and similarly average over all 55 locations. In all cases, the NPMM has lower spread, suggesting a better model fit. Finally, Table \ref{t:SVC_params} provides posterior means and SD of the GP parameters associated with the STVC model for the marginal parameters described in Section \ref{s:model_description}.
\bibliographystyle{elsarticle-num-names} 
\bibliography{SpatExtreme}





\end{document}

%% file: commands.tex
\newcommand{\btheta}{ \mbox{\boldmath $\theta$}}

\newcommand{\bx}{ \mbox{\bf x}}

\newcommand{\bs}{ \mbox{\bf s}}

\newcommand{\iid}{\stackrel{iid}{\sim}}

\newcommand{\calN}{{\cal N}}

\newcommand{\beq}{ \begin{equation}}
\newcommand{\eeq}{ \end{equation}}
\newcommand{\beqn}{ \begin{eqnarray}}
\newcommand{\eeqn}{ \end{eqnarray}}

\usepackage{caption}